\documentclass[pre,aps,twocolumn,superscriptaddress,floatfix]{revtex4}

\usepackage{graphicx}
\usepackage{amsmath,empheq}
\usepackage{amssymb}
\usepackage{ifthen}
\usepackage{booktabs}
\usepackage{bm}

\usepackage{graphicx}
\usepackage{dcolumn}
\usepackage{bm}
\usepackage{gensymb}
\usepackage{color}
\usepackage{gensymb}

\newcommand{\bb}{\begin{equation}}
\newcommand{\ee}{\end{equation}}
\newcommand{\ba}{\begin{eqnarray*}}
\newcommand{\ea}{\end{eqnarray*}}

\newcounter{subfigcount}
\newcounter{figcount}
\newcommand{\subfloat}[3]{%
{\ifnum\thefigure=\thefigcount\stepcounter{subfigcount}%
\else\setcounter{figcount}{\thefigure}\setcounter{subfigcount}{1}\fi%
}%
\noindent%
\begin{minipage}[b]{#1}%
  \centering%
  {#3}\\[0pt]%
  (\alph{subfigcount})~#2%
\end{minipage}}%

\newcommand{\subfloatflex}[2]{%
{\ifnum\thefigure=\thefigcount\stepcounter{subfigcount}
\else\setcounter{figcount}{\thefigure}\setcounter{subfigcount}{1}\fi%
}%
\noindent%
\begin{minipage}[b]{\widthof{#2}}%
  \centering%
  {#2}\\[0pt]%
  (\alph{subfigcount})~#1%
\end{minipage}}%

\makeatletter
\def\@frameeq#1{%
  \framebox{$\,\displaystyle#1\hbox{\vrule height 2.4ex depth 1.4ex width 0pt}\,$}}
\newcommand\Equation[1]{$$\refstepcounter{equation}%
  \@frameeq{#1}%
  \eqno \hbox{\@eqnnum}$$\@ignoretrue\ignorespaces}
\newcommand\Displaystyle[1]{$$\@frameeq{#1}$$\@ignoretrue\ignorespaces}
\makeatother

\begin{document}

\title{Capillary condensation between parallel walls of unequal length}

\author{Alexandr \surname{Malijevsk\'y}}
\affiliation{The Czech Academy of Sciences, Institute of Chemical Process Fundamentals,  Department of Molecular Modelling, 165 02 Prague, Czech
        Republic;}
\affiliation{Department of Physical Chemistry, University of Chemical Technology Prague, 166 28 Prague, Czech Republic}


\begin{abstract}
\noindent We present a macroscopic theory of capillary condensation in slits formed by parallel walls of unequal length. Using the concept of an edge
contact angle, we identify four distinct condensation states and derive Kelvin-like relations for their onset. The resulting phase diagrams,
expressed in terms of wall geometry and contact angle, reveal two central organizing features: a geometric separatrix that divides distinct
condensation regimes, and the wedge-filling threshold at $\theta=\pi/4$, which separates a rich four-state scenario from a simpler two-state one.
These results demonstrate how geometry dictates the onset and suppression of condensation in confined systems.
\end{abstract}

\maketitle

\section{Introduction}

The phase behaviour of confined fluids has long attracted considerable scientific interest, both for its fundamental implications and its
technological relevance. Early theoretical insights into the influence of confinement on fluid condensation date back to Lord Kelvin in the 19th
century, who showed that coexistence of capillary gas (CG) and capillary liquid (CL) in a macroscopically long slit of width $L$ requires a curved
meniscus of  Laplace radius $R$, associated with a pressure difference $\delta p=p_g-p_l=\gamma/R$, where $\gamma$ is the liquid-gas surface tension
\cite{thomson}.

Typically, the slit is assumed to be immersed in bulk gas, so equilibrium between CG and CL corresponds to capillary condensation -- the shift of
bulk vapour-liquid coexistence towards undersaturation. In terms of the chemical potential this shift is $\delta\mu_{cc}(T,L)\equiv\mu_{\rm
sat}(T)-\mu_{cc}(T,L)>0$, where $\mu_{\rm sat}(T)$  is the bulk saturation chemical potential at temperature $T$ and $\mu_{cc}((T,L)$ is the chemical
potential corresponding to capillary condensation. For large $L$, $\delta\mu_{cc}$ follows from expanding $\delta p$ around $\mu_{\rm sat}(T)$ to
first order. Using the geometric relation $R=L\sec\theta$, with $\theta$ the Young contact angle of the walls, one obtains
 \bb
 \delta\mu_{\rm cc}=\frac{2\gamma\cos\theta}{L\Delta\rho}\,, \label{kelvin}
 \ee
where $\Delta\rho=\rho_l-\rho_g$ is the density difference of the two bulk phases. This is the Kelvin equation for an infinitely long slit, although
it is often used in modified forms under additional assumptions, particularly at temperatures far below the critical point (see, e.g., \cite{gregg}).
From Eq.~(1) it is also clear that capillary condensation requires hydrophilic walls, $\theta<\pi/2$.

In the 1980s, capillary condensation was re-examined within the framework of modern statistical mechanics of simple fluids \cite{hansen,row}. Using
finite-size scaling arguments, Fisher and Nakanishi   showed that capillary condensation terminates at a capillary critical temperature $T_c(L)$
below the bulk critical temperature $T_c$ \cite{fisher81,nakanishi83}.  The  shift, $T_c - T_c(L) \propto L^{-\nu}$, where $\nu$ is the critical
exponent for the bulk correlation length, revealed a direct connection between confined and bulk criticality. They further argued that  confinement
not only shifts the critical point but also alters its nature, reflecting the effective reduction of dimensionality. Later, Evans and co-workers
\cite{evans84,evans86,evans87} elaborated the thermodynamics of confined fluids, formulated the corresponding Clapeyron equations, and interpreted
Kelvin's equation within the classical density functional theory (DFT) \cite{evans79}. These developments clarified how capillary condensation
relates to wetting \cite{dietrich,schick}, incorporated Derjaguin's correction \cite{derj,evans_marc85} and showing that the Kelvin equation can be
understood as the large-scale limit of microscopic theory. Subsequent DFT studies \cite{tar87,kierlik,neimark01}, simulations (see \cite{gelb} and
references therein), and experiments \cite{gamble81, fish_israel,hons10,zhong18,geim} confirmed that, despite its macroscopic origin, the Kelvin
equation provides a surprisingly accurate description of confinement even at nanoscales.

\begin{figure}[bht]
    \includegraphics[width=\linewidth]{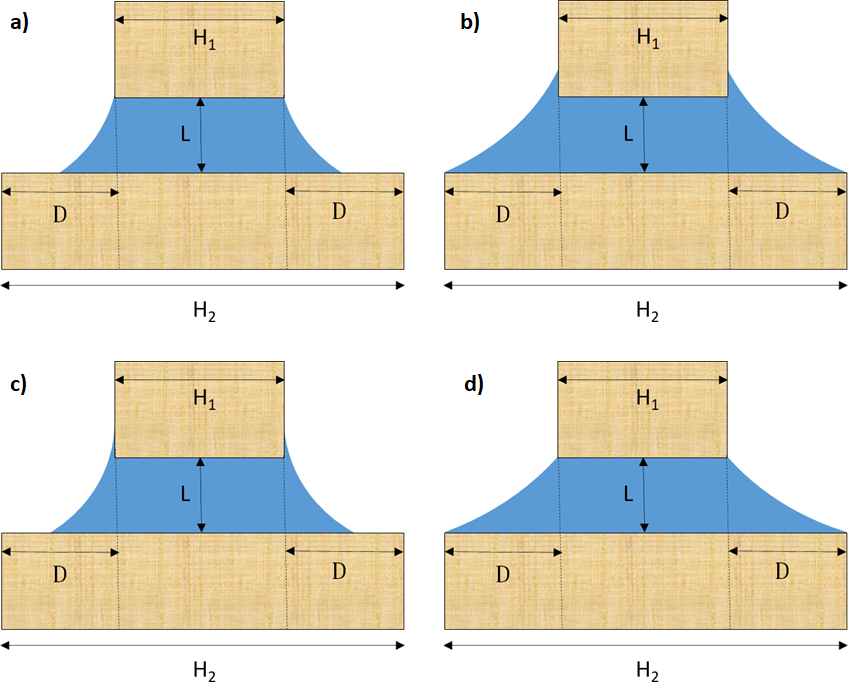}%
  \caption{Schematic representations of possible condensed states: a) $1^+$ state, b $1^-$ state, c) $0$ state, and d) $2$ state.  } \label{schemes}
\end{figure}

Beyond its fundamental interest, capillary condensation also plays a significant role in a variety of microscale systems. In microfluidic channels,
printing heads, and porous materials, small variations in the local geometry can determine where liquid first condenses and how capillary forces
develop. Similar phenomena occur in micro- and nano-mechanical structures, where the formation of liquid bridges may enhance adhesion or hinder
mechanical motion. Since real confining surfaces are often finite or exhibit abrupt terminations, it is important to understand how deviations from
the ideal infinite-slit geometry influence the onset and morphology of condensed phases.

These considerations naturally motivate the study of capillary condensation in finite slits, where edge effects and the limited extent of confining
walls play a central role. Hence, a natural extension of the classical slit model is to account for finite wall length $H$, so that edge effects
become important  \cite{finite_slit}. In this case, condensation is no longer governed by Young's contact angle $\theta$, but by an \emph{edge
contact angle} $\theta_e$, which characterizes how the menisci formed at the slit sides connect to the wall edges. The value of $\theta_e$, depending
on the ratio $H/L$, always exceeds $\theta$ and replaces it in Eq.~(\ref{kelvin}), thereby reducing the chemical potential shift $\delta\mu$ relative
to the infinite-slit case.

The present work generalizes capillary condensation theory to more realistic confinement geometries where one wall is truncated. Such partial
confinement is ubiquitous in practical situations. Micro- and nanofluidic channels often possess finite-length ceilings or overhangs due to
fabrication limitations, leading to regions where the confinement abruptly terminates.  Even in porous or granular media, the local confinement is
rarely perfectly symmetric, and fluid bridges frequently nucleate at truncated boundaries. By incorporating such geometrical features, the present
model captures the conditions under which technologically relevant capillary forces and condensation processes actually take place.

More specifically, we consider a slit model where the top wall (say) has length $H_1$, while the bottom wall has length $H_2>H_1$,  producing lateral
overhangs of size $D=(H_2-H_1)/2$. We still assume that the system is macroscopically deep and remains thus translationally invariant along the
walls. We show that this geometry admits multiple condensed states, distinguished by the location of the menisci. Unlike in equal-length finite
slits, where the menisci are uniquely pinned to the edges, this geometry permits several possible configurations (see Fig.~\ref{schemes}): the $1^+$
state, with menisci pinned to the top wall edges; the $1^-$ state, pinned to the bottom edges; the $0$ state, where the menisci meet both walls at
the equilibrium contact angle without pinning; and the $2$ state, where both walls pin the menisci.
%
%
%
%
%
Our aim is to classify these states, derive Kelvin-like equations for their onset, and analyze  the resulting phase behaviour. This analysis reveals
deep connections between capillarity, wetting, wedge-filling \cite{hauge92,rejmer99,parry99,mal13}, and pinning-depinning phenomena, and predicts
rich phase diagrams, with qualitative changes across the wedge-filling threshold $\theta=\pi/4$.

The remainder of this paper is organized as follows. Section II derives Kelvin-like equations for each condensation type, beginning with complete
wetting ($\theta=0$) and extending to partial wetting ($\theta>0$). Section III determines which condensation type occurs as a function of the
dimensionless parameters $\tilde{H}_1 = H_1/L$, $\tilde{D} = D/L$, and $\theta$, highlighting the connection with wedge-filling transitions. Section
IV examines the condensation limits and corresponding phase boundaries. Section V presents the global phase diagrams, with emphasis on the
qualitative changes across the wedge-filling threshold $\theta = \pi/4$. Finally, Section VI summarizes the main results and outlines possible
extensions.

\section{Condensation states and  Kelvin's equations}

In this section, we classify the possible condensed states and derive the corresponding Kelvin-like equations describing their onset. The analysis
proceeds in two steps: first, for completely wet walls ($\theta=0$), followed by a straightforward extension to partial wetting ($\theta>0$). This
distinction is important, since, the two regimes yield different phase behaviour, as shown below.

\subsection{Completely wet walls}

\subsubsection{$1^+$-condensation}

\begin{figure}[bht]
\includegraphics[width=\linewidth]{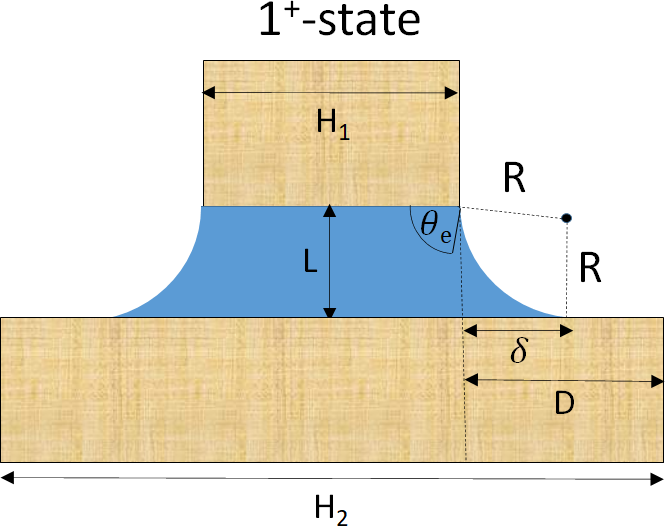}
\caption{Schematic of a $1^+$ state for completely wet walls. The meniscus of radius $R$ is pinned at the edges of the top wall, while meeting the
bottom wall tangentially. The edge contact angle at the top wall is $\theta_e$, and $\delta$ denotes the horizontal extension at which the meniscus
meets the bottom wall. } \label{scheme_1p_cw}
\end{figure}

We begin with the $1^+$ state, which is the most common condensation scenario. The condensed phase is separated from the surrounding gas by a pair of
cylindrical menisci of Laplace radius $R=\gamma/\delta p$, which meet the bottom wall tangentially, consistent with the equilibrium Young contact
angle $\theta=0$. By contrast, at the top wall the menisci meet the edges at the pressure-dependent edge contact angle $\theta_e$
(Fig.~\ref{scheme_1p_cw}), which is related to $R$ by
   \bb
   R(1+\cos\theta_e)=L\,.  \label{geom_1p_cw}
   \ee

The excess grand potential per unit length of the $1^+$ state, relative to the low-density (gas-like) state, is
 \bb
\Omega^{\rm ex}_{1^+}=\delta p(S+2\Delta S)+2\gamma\ell_m-2\gamma(H_1+\delta)\,. \label{om_1p_cw}
 \ee
The first term on the rhs corresponds to the free-energy cost of the metastable liquid; here $S=H_1L$ is the liquid area inside the slit, while
$\Delta S$ accounts for the liquid outside:
  \bb
   \Delta S=L\delta-\frac{R^2}{2}\left(\pi-\theta_e\right)-\frac{R^2}{2}\sin\theta_e\cos\theta_e\,.  \label{ds_1p_cw}
  \ee
The second term in Eq.~(\ref{om_1p_cw}) represents the free-energy cost due to the presence of the menisci, each of length $\ell_m$, and the last
term the wall-fluid contribution that favours the liquid phase. Finally, the outer extension $\delta$ is related to $\theta_e$ by
 \bb
\delta=R\sin\theta_e\,.  \label{delta_1p_cw}
 \ee

A first-order transition to the $1^+$ state occurs when $\Omega^{\rm ex}_{1^+}=0$. Substituting $R$ from Eq.~(\ref{geom_1p_cw}) yields the condition
  \bb
   \frac{\pi-\theta_e^{cc}+\sin\theta_e^{cc}\cos\theta_e^{cc}}{\sin^2\theta_e^{cc}}=\frac{H_1}{L}\,,  \label{thetae_1p_cw}
  \ee
which determines the condensation edge angle, $\theta_e^{cc}$, as a function of $H_1/L$. This equation does not involve $D$ but assumes implicitly
that $D>\delta$. To ensure this, $\theta_e^{cc}$ must also satisfy
    \bb
   \tan\left(\frac{\theta_e^{cc}}{2}\right)<\frac{D}{L}\,, \label{cond_1p_cw}
    \ee
as implied by Eqs.~(\ref{geom_1p_cw}) and (\ref{delta_1p_cw}).

The Kelvin-like equation for $1^+$ condensation then follows as
  \bb
   \delta \mu_{cc}^{(1^+)}=\frac{\gamma(1+\cos\theta_e^{cc})}{\Delta\rho L}\,, \label{kelvin_1p_cw}
  \ee
with $\theta_e^{cc}$ determined by Eq.~(\ref{thetae_1p_cw}).

\subsubsection{$1^-$-condensation}

\begin{figure}[bht]
\includegraphics[width=0.8\linewidth]{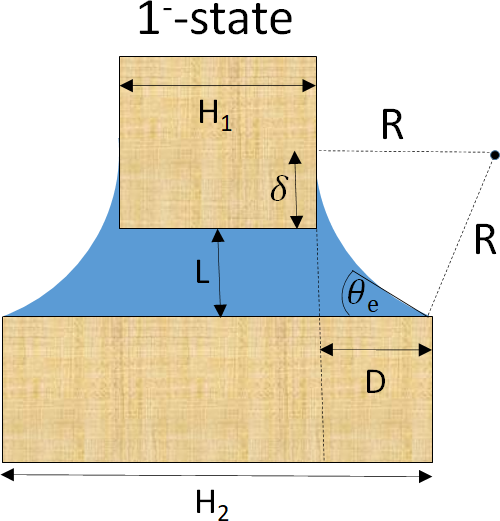}
\caption{Schematic of a $1^-$ state for completely wet walls. The meniscus of radius $R$ is pinned at the edges of the bottom wall, while meeting the
top wall tangentially. The edge contact angle at the bottom wall is $\theta_e$, and $\delta$ denotes the vertical extension at which the meniscus
meets the top wall. } \label{scheme_1m_cw}
\end{figure}

The $1^-$ state mirrors the $1^+$ configuration, with menisci now pinned at the bottom wall edges and meeting the top wall tangentially
(Fig.~\ref{scheme_1m_cw}). The geometry requires
  \bb
  R(1-\sin\theta_e)=D\,, \label{geom_1m_cw}
  \ee
with extension
 \bb
 \delta=R\cos\theta_e-L \,.\label{delta_1m_cw}
 \ee

The excess grand potential is
  \bb
 \frac{\Omega^{\rm ex}_{1^-}}{\gamma}=\frac{H_1L+2\Delta S}{R}+2\ell_m-2(H_1+D+\delta)\,, \label{om_1m_cw}
 \ee
 with
 \bb
 \Delta S=R^2\cos\theta_e-\frac{R^2}{2}\left(\frac{\pi}{2}-\theta_e\right)-\frac{R^2}{2}\sin\theta_e\cos\theta_e\,,   \label{ds_1m_cw}
 \ee
 and
 \bb
 \ell_m=R\left(\frac{\pi}{2}-\theta_e\right)\,. \label{ell_m}
 \ee
Imposing $\Omega^{\rm ex}_{1^-}=0$ yields
 \begin{eqnarray}
 &&H_1L(1-\sin\theta_e^{cc})^2+D^2\left(\frac{\pi}{2}-\theta_e^{cc}-\sin\theta_e^{cc}\cos\theta_e^{cc}\right)=\nonumber\\
 &&=2D(H_1+D+L)\,,  \label{thetae_1m_cw}
 \end{eqnarray}
which fixes the condensation edge angle $\theta_e^{cc}$ for given geometry. To ensure that the menisci connect tangentially to the top wall
($\delta>0$), $\theta_e^{cc}$ must also satisfy
  \bb
 \tan\left(\frac{\pi}{4}-\frac{\theta_e}{2}\right)<\frac{D}{L}\,, \label{cond_1m_cw}
  \ee
 as implied by Eqs.~(\ref{geom_1m_cw}) and (\ref{delta_1m_cw}).

Once $\theta_e^{cc}$ is determined, the corresponding Kelvin-like equation for the $1^-$ state follows as
 \bb
 \delta \mu_{cc}^{(1^-)}=\frac{\gamma(1-\sin\theta_e^{cc})}{\Delta\rho D}\,. \label{kelvin_1m_cw}
 \ee

\subsubsection{$2$-condensation}

\begin{figure}[bht]
\includegraphics[width=0.8\linewidth]{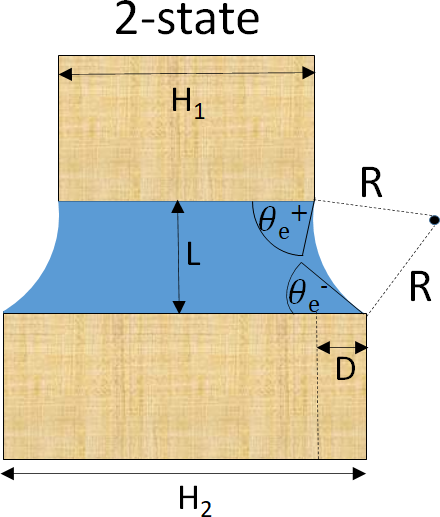}
\caption{Schematic of a $2$ state for completely wet walls. The meniscus of radius $R$ is simultaneously pinned at both walls, with edge contact
angles $\theta_e^+$ (top) and $\theta_e^-$ (bottom). } \label{scheme_2_cw}
\end{figure}

The $2$ state is characterized by simultaneous pinning of the menisci at both walls, with edge contact angles $\theta_e^+$ and $\theta_e^-$
(Fig.~\ref{scheme_2_cw}). The geometry requires
 \bb
  R(\cos\theta_e^++\cos\theta_e^-)=L \label{geom1_2_cw}
 \ee
 and
 \bb
 R(\sin\theta_e^+-\sin\theta_e^-)=D\,, \label{geom2_2_cw}
 \ee
implying
 \bb
  \tan\left(\frac{\theta_e^{+}-\theta_e^{-}}{2}\right)=\frac{D}{L}\,. \label{geom_2_cw}
 \ee


The excess grand potential is
  \bb
 \frac{\Omega^{\rm ex}_{2}}{\gamma}=\frac{H_1L+2\Delta S}{R}+2\ell_m-2(H_1+D)\,, \label{om_2_cw}
 \ee
with
 \begin{eqnarray}
  \Delta S&=&LR\sin\theta_e^+-\frac{R^2}{2}\left(\pi-\theta_e^+-\theta_e^-\right)\nonumber\\
  &&-R^2(\sin\theta_e^+\cos\theta_e^++\sin\theta_e^-\cos\theta_e^-) \label{ds_2_cw}
 \end{eqnarray}
 and
 \bb
 \ell_m=R(\pi-\theta_e^+-\theta_e^-)\,. \label{lm_2_cw}
 \ee

Condensation into the $2$ state occurs when the upper and lower edge contact angles take the values $\theta_e^{+cc}$ and $\theta_e^{-cc}$,
respectively. Imposing $\Omega^{\rm ex}_{2}=0$ leads to
 \begin{eqnarray}\label{thetae_2_cw}
 &&\!\!\! \!\!\!\!\!\!\!\!H_1L(\sin\theta_e^{+cc}-\sin\theta_e^{-cc})^2
+2LD\sin\theta_e^{+cc}(\sin\theta_e^{+cc}-\sin\theta_e^{-cc})\nonumber\\
&&\!\!\! \!\!\!\!\!\!\!+D^2(\pi-\theta_e^{+cc} -\theta_e^{-cc}
-\sin\theta_e^{+cc}\cos\theta_e^{+cc}+\sin\theta_e^{-cc}\cos\theta_e^{-cc})\nonumber\\
&&\!\!\! \!\!\!\!\!\!\!=2(\sin\theta_e^{+cc}-\sin\theta_e^{-cc})D(H_1+D)\,.
 \end{eqnarray}

Together with Eq.~(\ref{geom_2_cw}), this relation determines the condensation angles $\theta_e^{\pm cc}$, which in turn specify the transition
location via the Kelvin-like equation for 2-condensation:
 \bb
\delta\mu_{cc}^{(2)}=\frac{\gamma(\cos\theta_e^{+cc}+\cos\theta_e^{-cc})}{\Delta\rho L}\,, \label{kelvin_2_cw}
 \ee
where Eq.~(\ref{geom1_2_cw}) has been used.

\subsubsection{$0$-condensation}

\begin{figure}[bht]
\includegraphics[width=\linewidth]{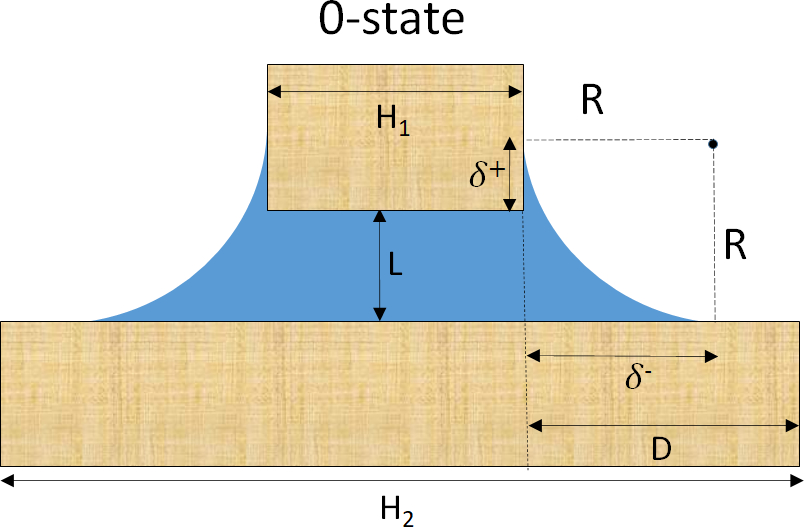}
\caption{Schematic of a $0$ state for completely wet walls. The menisci of radius $R$ meet both walls tangentially, without being pinned at edges.
The extensions are $\delta^+=R-L$ (top) and $\delta^-=R$ (bottom).} \label{scheme_0_cw}
\end{figure}

The $0$ state is qualitatively distinct: no edge contact angle appears, as the menisci meet both walls at the equilibrium Young contact angle
(Fig.~\ref{scheme_0_cw}).

The excess grand potential  is
  \bb
    \frac{\Omega^{\rm ex}_{0}}{\gamma}=\frac{LH_1}{R}+\left(\frac{\pi}{2}-2\right)R-2(H_1-L)\,,\label{om_0_cw}
  \ee
which vanishes at condensation when $R=R_{cc}$. The condensation radius is given by
   \bb
  R_{cc}=\frac{\sqrt{(H_1-L)^2+\left(2-\frac{\pi}{2}\right)LH_1}-(H_1-L)}{2-\frac{\pi}{2}} \label{Rcc_0_cw}
  \ee
and yields the Kelvin-like equation for 0 condensation:
 \bb
   \delta \mu_{cc}^{(0)}=\frac{\gamma}{\Delta\rho R_{cc}}\,.  \label{kelvin_0_cw}
   \ee

In addition, the geometric constraints $R>L$ and $R<D$ imply
   \bb
   \frac{2LD-\left(2-\frac{\pi}{2}\right)D^2}{2D-L}<H_1<\frac{\pi}{2}L\,,  \label{cond_0_cw}
   \ee
 with $D>L$ required by construction.

\subsection{Partially wet walls}

In the following, we generalize the results from the previous paragraph by considering $\theta>0$.

\subsubsection{$1^+$-condensation}

\begin{figure}[bht]
\includegraphics[width=\linewidth]{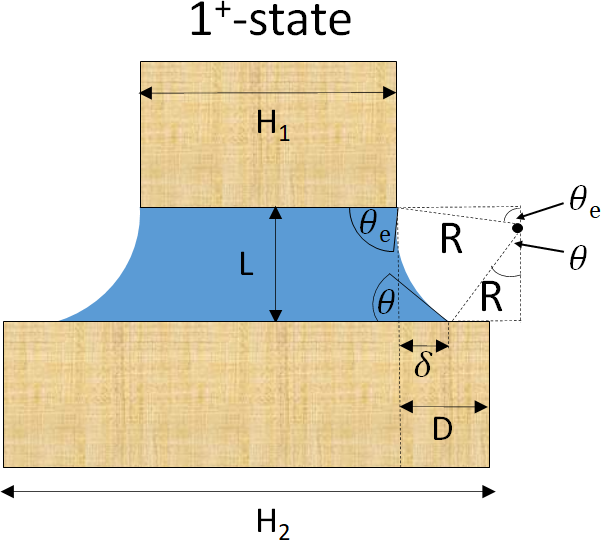}
\caption{Schematic of a $1^+$ state for partially wet walls. The meniscus of radius $R$ meets the bottom wall at Young's contact angle $\theta$,
while at the top wall it is pinned at the edges with edge contact angle $\theta_e$. The horizontal extension $\delta$ denotes where the meniscus
intersects the bottom wall.} \label{scheme_1p_pw}
\end{figure}

For partially wet walls, the geometry of the $1^+$ state dictates (see Fig.~\ref{scheme_1p_pw})
 \bb
 R(\cos\theta+\cos\theta_e)=L \label{geom_1p_pw}
 \ee
 and
 \bb
 \delta=R(\sin\theta_e-\sin\theta)\,,  \label{delta_1p_pw}
 \ee
  generalizing Eqs.~(\ref{geom_1p_cw}) and (\ref{delta_1p_cw}), respectively.

 \vspace*{0.5cm}

At condensation, the free-energy balance yields
 \begin{eqnarray}
 &&\frac{H_1L}{R_{cc}}+2L\sin\theta_e^{cc}-R_{cc}(\cos\theta_e^{cc}\sin\theta_e^{cc}+\cos\theta\sin\theta)\nonumber\\
 &&+R_{cc}(\pi-\theta-\theta_e^{cc})=2(H_1+\delta)\cos\theta\,,  \label{fe_1p}
 \end{eqnarray}
which, after substituting from Eqs.~(\ref{geom_1p_pw}) and (\ref{delta_1p_pw}), reduces to
 \begin{eqnarray}
 &&\tilde{H}_1(\cos^2\theta_e^{cc}-\cos^2\theta)+\sin\theta_e^{cc}(2\cos\theta+\cos\theta_e^{cc})\nonumber\\
 &&+\pi-\theta-\theta_e^{cc}=\cos\theta\sin\theta\,,  \label{thetae_1p_pw}
 \end{eqnarray}
 where  recall $\tilde{H_1}=H_1/L$.

The solution of Eq.~(\ref{thetae_1p_pw}) determines the condensation edge angle $\theta_e^{cc}$, which fixes the transition location via the
Kelvin-like equation
  \bb
   \delta \mu_{cc}^{(1^+)}=\frac{\gamma(\cos\theta+\cos\theta_e^{cc})}{\Delta\rho L}\,. \label{kelvin_1p_pw}
  \ee

Finally, the geometric constraint $D>\delta$ requires
   \bb
   \tan\left(\frac{\theta_e-\theta}{2}\right)<\frac{D}{L}\,, \label{cond_1p_pw}
   \ee
generalizing the condition (\ref{cond_1p_cw}) obtained for completely wet walls.

\subsubsection{$1^-$-condensation}

\begin{figure}[bht]
\centerline{\includegraphics[width=0.9\linewidth]{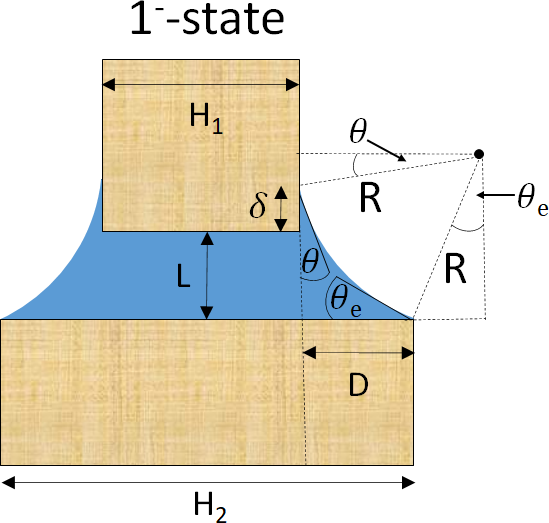}}
 \caption{Schematic of a $1^-$ state for partially wet walls. The meniscus of radius $R$ meets the top wall at Young's contact angle $\theta$, while at the
bottom wall it is pinned with edge contact angle $\theta_e$. The vertical extension $\delta$ measures the intersection with the top wall.}
\label{scheme_1m_pw}
\end{figure}

In the partially wet $1^-$ state, the menisci meet the top wall at the intrinsic Young contact angle $\theta$ a distance $\delta$ from the edges,
while the bottom wall is met at its edges with the pressure-dependent edge contact angle $\theta_e$ (Fig.~\ref{scheme_1m_pw}). The geometry implies
the relations
 \bb
 R(\cos\theta-\sin\theta_e)=D\,,  \label{geom_1m_pw}
 \ee
and
 \bb
 \delta=R(\cos\theta_e-\sin\theta)-L\,, \label{delta_1m_pw}
 \ee
with the additional condition
 \bb
 \tan\left(\frac{\pi}{4}+\frac{\theta-\theta_e}{2}\right)<\frac{D}{L}\,. \label{cond_1m_pw}
  \ee

The excess grand potential per unit length is
 \begin{eqnarray}
  \frac{\Omega^{\rm ex}_{1^-}}{\gamma}&=&\frac{H_1L}{R}+2R\cos\theta\cos\theta_e+R\left(\frac{\pi}{2}-\theta-\theta_e\right)\nonumber\\
  &&-R(\sin\theta_e\cos\theta_e+\sin\theta\cos\theta)\\
  &&-2(H_1+D+R\cos\theta_e-R\sin\theta-L)\cos\theta\,,\nonumber \label{om_1m_pw}
 \end{eqnarray}
which vanishes at condensation when the edge contact angle takes the value $\theta_e^{cc}$. Combining $\Omega^{\rm ex}_{1^-}=0$ with
Eq.~(\ref{geom_1m_pw}) yields
 \begin{widetext}
   \bb
 H_1L(\cos\theta-\sin\theta_e^{cc})^2+D^2
 \left(\frac{\pi}{2}-\theta-\theta_e^{cc}+\sin\theta\cos\theta-\sin\theta_e^{cc}\cos\theta_e^{cc}\right)
 =2D(H_1+D-L)\cos\theta(\cos\theta-\sin\theta_e^{cc})\,. \label{thetae_1m_pw}
 \ee
 \end{widetext}

Eq.~(\ref{thetae_1m_pw}) determines $\theta_e^{cc}$, which in turn yields the location of the transition via the Kelvin-like equation
 \bb
 \delta \mu_{cc}^{(1^-)}=\frac{\gamma(\cos\theta-\sin\theta_e^{cc})}{\Delta\rho D}\,. \label{kelvin_1m_pw}
 \ee


\subsubsection{$2$-condensation}

\begin{figure}[t]
\includegraphics[width=0.9\linewidth]{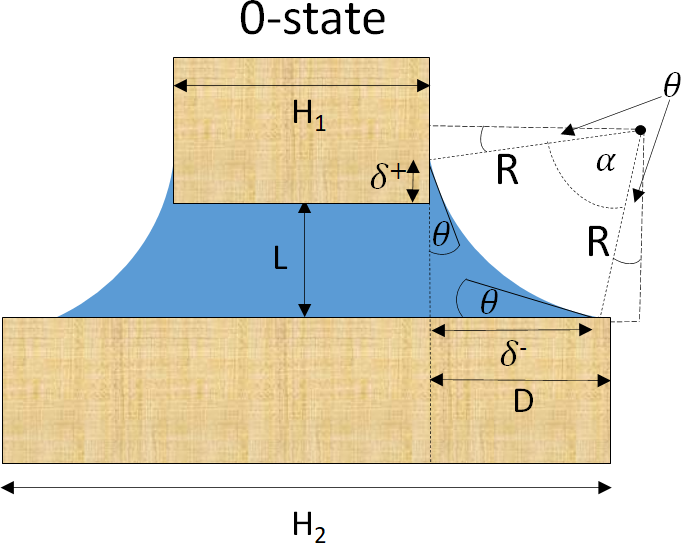}
\caption{Schematic of a $0$ state for partially wet walls. The menisci of radius $R$ meet both walls at the intrinsic angle $\theta$, with extensions
$\delta^-$ (bottom wall) and $\delta^+$ (top wall).} \label{scheme_0_pw}
\end{figure}

The $2$ state for partially wet walls is geometrically the same as in the completely wet case: menisci are pinned at both walls, with edge contact
angles $\theta_e^+$ and $\theta_e^-$ subject to relations (\ref{geom1_2_cw})--(\ref{geom_2_cw}). The only modification arises in the wall-fluid
free-energy term of Eq.~(\ref{om_2_cw}), which now has a factor of $\cos\theta$.

As a result, the condition $\Omega^{\rm ex}_{2}=0$ becomes
  \begin{eqnarray}
  &&\frac{H_1L}{R_{cc}}+2L\sin\theta_e^{+cc}+R_{cc}(\pi-\theta_e^{+cc}-\theta_e^{-cc})\nonumber\\
  &&-R(\sin\theta_e^{+cc}\cos\theta_e^{+cc}+\sin\theta_e^{-cc}\cos\theta_e^{-cc})\nonumber\\
  &&=2\cos\theta(H_1+D)\,, \label{theta_two_pw}
   \end{eqnarray}
which alters the equilibrium values of the edge contact angles $\theta_e^{+cc}$ and $\theta_e^{-cc}$ that appear in the Kelvin equation
(\ref{kelvin_2_cw}).

\subsubsection{$0$-condensation}

In the $0$ state with partially wet walls, the menisci meet both walls at the intrinsic Young contact angle $\theta$. The corresponding excess grand
potential  is
 \begin{eqnarray}  \label{om_0_pw}
   \frac{\Omega^{\rm ex}_{0}}{\gamma}&=&
   \frac{H_1L}{R_{cc}}+2R_{cc}\cos\theta(\cos\theta-\sin\theta)\\
   &&+R_{cc}\left(\frac{\pi}{2}-2\theta\right)-2(H_1+\delta^++\delta^-)\cos\theta\nonumber\,,
 \end{eqnarray}
 where the extensions $\delta^-$ and $\delta^+$ are
  \bb
 \delta^-=R(\cos\theta-\sin\theta)\,, \label{g0a}
 \ee
 and
\bb
 \delta^+=\delta^--L\,. \label{g0b}
\ee

Substituting Eqs.~(\ref{g0a}) and (\ref{g0b}) into Eq.~(\ref{om_0_pw}) and imposing $\Omega^{\rm ex}_{0}=0$ leads to
 \begin{eqnarray}
  &&\left(\frac{\pi}{4}-\theta-\cos^2\theta+\cos\theta\sin\theta\right)R_{cc}^2\nonumber\\
  &&-\cos\theta(H_1-L)R_{cc}+\frac{H_1L}{2}=0\,, \label{R0}
 \end{eqnarray}
which determines the condensation radius:
 \begin{widetext}
 \bb
  R_{cc}=\frac{\cos\theta(H_1-L)+\sqrt{\cos^2\theta(H_1-L)^2-H_1L\left(\frac{\pi}{2}-2\theta-2\cos^2\theta+2\cos\theta\sin\theta\right)}}
  {\frac{\pi}{2}-2\theta-2\cos^2\theta+2\cos\theta\sin\theta} \label{Rcc_0_pw}
 \ee
 \end{widetext}
and thus the location of the transition via
 \bb
 \delta \mu_{cc}^{(0)}=\frac{\gamma}{\Delta\rho R_{cc}}\,.
 \ee

Stability of the $0$ state further requires
 \bb
  R(\cos\theta-\sin\theta)>L\;\;\;{\rm and}\;\;\;R(\cos\theta-\sin\theta)<D\,,
 \ee
which combine into
 \bb
  \frac{aLD+\left(\frac{\pi}{2}-2\theta-a\right)D^2}{aD-(\cos\theta-\sin\theta)^2L}<H_1<\frac{(\pi-4\theta)L}{4\cos^2\theta-2}\,,   \label{cond_0_pw}
 \ee
 with $a=2\cos\theta(\cos\theta-\sin\theta)$.

\section{Phase boundaries between condensed states}

In this section, we determine the phase boundaries separating the condensed states and analyze their properties. We begin by formulating two
observations:\\

 \noindent{\bf Observation 1:}\\
 \noindent\emph{For any $\theta\le\pi/4$, there exists a unique point at $\tilde{D}=1$ and
   \bb
 \tilde{H}_1=\frac{\pi-4\theta}{4\cos^2\theta-2}\,, \label{4coex}
  \ee
ranging from $\pi/2$ (for $\theta=0$) to $1$ (for $\theta=\pi/4$), for which all four condensed states collapse into a single state that coexists
with the gas phase. The corresponding undersaturation is
 \bb
 \delta\mu_{\rm cc}=\frac{2\gamma(\cos\theta-\sin\theta)}{L\Delta\rho}\,.\\ \label{kelvin2}
 \ee
 }

 \noindent These results follows directly from the Kelvin-like equations presented in the previous section.\\

 \noindent{\bf Observation 2:}\\
\noindent\emph{For $\theta\le\pi/4$, the line $\tilde{D}=1$ acts as a separatrix between $0$- and $2$-condensation,
  the only locus where the two can coexist. Specifically:\\
     (i) $0$-condensation exists if and only if $\tilde{D}\geq 1$\,;\\
    (ii) $2$-condensation exists if and only if $\tilde{D}\leq 1$\,.\\}

\begin{figure}[bht]
\includegraphics[width=0.9\linewidth]{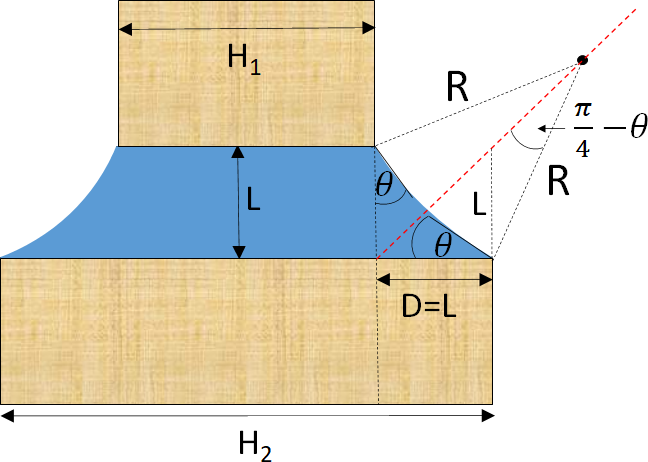}
\caption{Schematic illustrating the coexistence of the 0 and 2 states, which occurs only when $D = L$. For a given Young contact angle $\theta$, the
meniscus radius of curvature $R$ is determined by Eq.~(\ref{r02}), as implied by the construction, and is centered on the bisector forming a
$45^\circ$ angle with the bottom wall (red dashed line).} \label{20_coex}
\end{figure}

The first part of this observation follows directly from Eq.~(\ref{geom_2_cw}), noting that coexistence between $2$- and $0$-condensation requires
the edge contact angles $\theta_e^+=\pi/2+\theta$ and $\theta_e^-=\theta$. The geometry admits a simple interpretation, especially for $\theta=0$:
the coexistence state corresponds to menisci forming quarter-circle arcs of radius $R=D=L$. For $0\leq\theta\leq\pi/4$, the circle centers shift away
from the walls along the bisector of the slit corner, i.e. a line inclined at $45^\circ$ to the bottom walls, which enforces $D=L$ (see
Fig.~\ref{20_coex}). The corresponding meniscus radius is
 \bb
 R=\frac{L}{\sqrt{2}\sin\left(\tfrac{\pi}{4}-\theta\right)}\,, \label{r02}
  \ee
which diverges at $\theta=\pi/4$, beyond which the construction ceases to exist.

By construction, $0$-condensation requires $\tilde{D}\geq 1$. For such extensions, $0$-condensation can always be stabilized by adjusting
$\tilde{H}_1$. This is evident for large $\tilde{D}$ in the absence of $1^-$-condensation (see below), where sufficiently small $\tilde{H}_1$ yields
a large condensation radius of curvature. For smaller $\tilde{D}$ (but still greater than unity), the existence of $0$-condensation follows from the
fact that $1^+$ and $1^-$ states cannot coexist except at $\tilde{D}=1$, where all condensation states meet. Thus, by tuning $\tilde{H}_1$,
$0$-condensation necessarily fills the gap between the $1^+$ and $1^-$ regions.

In contrast, $2$-condensation is restricted to $\tilde{D}\leq 1$. More precisely, $\tilde{D}<1$ is not only a necessary but also a \emph{sufficient}
condition for the existence of $2$-condensation within some interval of $\tilde{H}_1$. Indeed, for any $\tilde{D}>0$, $1^+$-condensation can always
be stabilized at sufficiently large $\tilde{H}_1$. As $\tilde{H}_1$ is decreased at fixed $\tilde{D}$, the condensation menisci expand. With
$0$-condensation excluded in this regime, the menisci must inevitably become pinned at the bottom wall before detaching from the top wall, thereby
stabilizing $2$-condensation.\\

In the following subsections, we specify the explicit forms of the phase boundaries. As before, we first consider completely wet walls and then
extend the discussion to partial wetting, highlighting the contrasts between the two cases.

\subsection{Completely wet walls}

For $\theta=0$, the determination of phase boundaries is particularly transparent. As established above, the line $\tilde{D}=1$ acts as a separatrix,
with two distinct sets of phase boundaries emerging on either side:

\begin{enumerate}
 \item $\mathbf{D>L}$: The $0$-$1^-$ boundary follows by setting $R_{cc}=D$ in Eq.~(\ref{Rcc_0_cw}),
 or equivalently $\theta_e^{cc}=0$ in Eq.~(\ref{thetae_1m_cw}):
 \bb
  \tilde{H}_1^{0-1^-}=\frac{\left(\frac{\pi}{2}-2\right)\tilde{D}^2+2\tilde{D}}{2\tilde{D}-1}\,. \label{bound_01m_cw}
  \ee
  For large $\tilde{D}$ this decays linearly, $\tilde{H}_1^{0-1^-}\sim-(1-\pi/4)\tilde{D}$, while in the limit of
   $\tilde{D}\to1^+$ it approaches $\pi/2$ according to
   \bb
  \tilde{H}_1^{0-1^-}\sim \pi/2-2(\tilde{D}-1)\,,  \;\;\;\tilde{D}\to1^+\,.
    \ee

The $0$-$1^+$ boundary is instead obtained by imposing $R_{cc}=L$ in Eq.~(\ref{Rcc_0_cw}), or equivalently $\theta_e^{cc}=\pi/2$ in
Eq.~(\ref{thetae_1p_cw}), which yields the constant line
  \bb
  \tilde{H}_1^{0-1^+}=\frac{\pi}{2}\,.  \label{bound_01p_cw}
 \ee

Thus, for $D>L$, the $0$-$1^+$ boundary always lies above the $0$-$1^-$ line, and both meet at $D=L$, below which the $0$ state disappears.

 \item $\mathbf{D<L}$: In this regime, the $0$ state is absent but the system may instead undergo condensation into the $2$ state.

The $2$-$1^+$ boundary is reached when the lower edge contact angle of the 2-state, $\theta_e^{-cc}$, vanishes. From Eq.~(\ref{geom_2_cw}) this
implies
 \bb
 \tan\left(\frac{\theta_e^{+cc}}{2}\right)=\tilde{D}\,. \label{tan_half}
 \ee
Recognizing  the familiar  ``tangent half-angle substitution'', one substitutes   $\sin\theta_e^{+cc}=2\tilde{D}/(1+\tilde{D}^2)$,
$\cos\theta_e^{+cc}=(1-\tilde{D}^2)/(1+\tilde{D}^2)$ and $\theta_e^{+cc}=2\tan^{-1}\tilde{D}$  into Eq.~(\ref{thetae_2_cw}) to get
  \bb
\tilde{H}_1^{2-1^+}=\frac{(1+\tilde{D}^2)^2(\pi-2\tan^{-1}\tilde{D}) +2\tilde{D}(1-\tilde{D}^2)}{4 \tilde{D}^{2}}\,, \label{bound_21p_cw}
 \ee
 which defines the $2$-$1^+$ boundary.

As $\tilde{D}\to0$, this diverges as
  \bb
  \tilde{H}_1^{2-1^+}\sim\frac{\pi}{4}\tilde{D}^{-2}\,,\;\;\;\tilde{D}\to0\,, \label{2as_left_cw}
 \ee
 reflecting $\theta_e^{+cc}\ll1$ with $\theta_e^{+cc}\approx 2\tilde{D}+{\cal{O}}(D^3)$.

In the opposite limit, $\tilde{D}\to1$, one finds
\begin{equation}
\theta_e^{+cc}=\frac{\pi}{2}+1-\tilde{D}^{-1}\,,\;\;\; \tilde{H}_1^{2-1^+}\sim\frac{\pi}{2}+2(1-\tilde{D})\,, \label{2as_right_cw}
\end{equation}
showing linear approach to the four-state coexistence point at $D=L$.

The $\tilde{H}_1^{2-1^-}$ boundary arises when the upper edge angle satisfies $\theta_e^{+cc}=\pi/2$, which implies
 \bb
 \frac{1-\sin\theta_e^{-cc}}{\cos\theta_e^{-cc}}=\frac{D}{L}\,. \label{2c}
 \ee
The corresponding free-energy condition reads
 \begin{eqnarray}
 &&H_1^{2-1^-}\cos\theta_e^{-cc}+2L+\left(\frac{\pi}{2}-\theta_e^{-cc}\right)L\sec\theta_e^{-cc}\nonumber\\
 &&-\sin\theta_e^{-cc}=2(H_1+D)\,. \label{bound_21m_cw}
 \end{eqnarray}
Using Eq.~(\ref{2c}), one obtains
  \bb
\cos\theta_e^{-cc}=2\tilde{D}/(\tilde{D}^2+1)\equiv c\,,
 \ee
which allows the boundary to be expressed explicitly as
   \bb
\tilde{H}_1^{2-1^-}=\frac{\cos^{-1}c-\frac{\pi}{2}+2c(\tilde{D}-1)+c\sqrt{1-c^2}}{(c-2)c}\,. \label{bound_21m_cw2}
 \ee

Unlike the $2$-$1^+$ line, this boundary remains finite across the entire range of $\tilde{D}$. As $\tilde{D}\to0$, it approaches unity according to
  \bb
  \tilde{H}_1^{2-1^-}\sim 1+\frac{4}{3}\tilde{D}^2\,,\;\;\;\tilde{D}\to0\,, \label{21m_left_cw}
 \ee
 with $\theta_e^{-cc}\sim \pi/2-2\tilde{D}$. 

 In the opposite limit, $\tilde{D}\to1$, the boundary approaches $\pi/2$ also quadratically,
 \bb
  \tilde{H}_1\sim\frac{\pi}{2}-(\tilde{D}-1)^2\,,\;\;\;\tilde{D}\to1\,, \label{21m_right_cw}
 \ee
while the lower edge angle vanishes as $\theta_e^{-cc}\sim1-\tilde{D}$.

\end{enumerate}

\subsection{Partially wet walls}

For partially wet walls, the range of admissible condensation scenarios becomes more restricted. To construct the global phase diagrams, we identify
the loci of three-phase coexistence that mark the boundaries between different condensation types. We begin with the $0$-condensation region
(relevant only for $\theta<\pi/4$), and then determine the $2$-condensation boundaries, which together complete the overall picture.



\subsubsection{$0$-$1^-$ boundary}

0-condensation merges with $1^-$-condensation when the lower extension $\delta^-$, given by Eq.~(\ref{g0a}), just reaches the bottom wall edge
$\delta^-=D$. At this point, the Laplace radius is
 \bb
  R_{cc}=\frac{D}{\cos\theta-\sin\theta}\,, \label{R01m}
 \ee
which upon substitution into Eq.~(\ref{R0}), yields
 \begin{widetext}
\bb
 \tilde{H}_1^{0-1^-}=\frac{\tilde{D}^2\left(\frac{\pi}{2}-2\theta-2\cos^2\theta+2\cos\theta\sin\theta\right)+2\tilde{D}\cos\theta(\cos\theta-\sin\theta)}
 {2\tilde{D}\cos\theta(\cos\theta-\sin\theta)-(\cos\theta-\sin\theta)^2} \,,\;\;\;\left(\theta<\frac{\pi}{4}\right)\,, \label{bound_01m}
 \ee
 \end{widetext}
defining the lower boundary of the $0$-condensation region in the $\tilde{D}$-$\tilde{H}_1$ plane.

\subsubsection{$0$-$1^+$ boundary}

Here, the coexistence requires the upper extension $\delta^+$ from  Eq.~(\ref{g0b}) to vanish, corresponding to
 \bb
  R_{cc}=\frac{L}{\cos\theta-\sin\theta}\,. \label{R01p}
 \ee
Substituted this into Eq.~(\ref{R0}) yields the $\tilde{D}$-independent upper bound of the $0$-region [cf. Eq.~(\ref{4coex}],
  \bb
 \tilde{H}_1^{0-1^+}=\frac{\pi-4\theta}{4\cos^2\theta-2}\,, \label{bound_01p}
  \ee
  valid for  $\tilde{D}\ge1$ and providing the upper bound of the $0$-condensation region. One can readily
verify that $\tilde{H}_1^{0-1^+}$ and $\tilde{H}_1^{0-1^-}$ boundaries indeed meet at $\tilde{D}=1$.

\subsubsection{$2$-$1^+$ boundary}

To locate the $2$-$1^+$ boundary, we consider the $1^+$ state and require $\delta=D$. Substituting this into Eq.~(\ref{delta_1p_pw}) gives
 \bb
 R(\sin\theta_e-\sin\theta)=D\,, \label{delta_21}
 \ee
which, together with Eq.~(\ref{geom_1p_pw}), leads to
 \bb
 \frac{\sin\theta_e-\sin\theta}{\cos\theta+\cos\theta_e}=\frac{D}{L}\,. \label{g21p}
 \ee
 Next, substituting $R$ from Eq.~(\ref{geom_1p_pw}) into the free-energy balance (\ref{fe_1p}), with $\delta=D$, yields at the  $2$-$1^+$ boundary
 \begin{eqnarray}
 &&\tilde{H}_1^{2-1^+}(\cos^2\theta_e-\cos^2\theta)+\sin\theta_e\cos\theta_e-\sin\theta\cos\theta\label{fe21p}\\
 &&+2\sin\theta_e\cos\theta+\pi-\theta-\theta_e
 =2\tilde{D}\cos\theta(\cos\theta+\cos\theta_e)\nonumber\,.
 \end{eqnarray}
From Eqs.~(\ref{g21p}) and (\ref{fe21p}), one obtains an explicit expression for the location of the boundary:
 \begin{widetext}
 \bb
 \tilde{H}_1^{2-1^+}=\frac{\sin\theta\cos\theta+\cos^{-1}c+\theta-\pi-\sqrt{1-c^2}(2\cos\theta+c)+2\tilde{D}\cos\theta(\cos\theta+c)}{c^2-\cos^2\theta}
  \label{bound_21p}
 \ee
 \end{widetext}
where $c\equiv\cos\theta_e$ is determined from Eq.~(\ref{g21p}) as
  \bb
 c=\frac{\cos\theta- \tilde{D}^2\cos\theta -2 \tilde{D} \sin\theta }{\tilde{D}^2+1}\,.
  \ee

The same result arises by considering the $2$-state with $\theta_e^-=\theta$ in Eq.~(\ref{theta_two_pw}). Note that $\tilde{H}_1^{2-1^+}$ retains a
single pole at $\tilde{D}=0$.


We now examine the asymptotic behaviour of $\tilde{H}_1^{2-1^+}$:

\begin{enumerate}

\item  $\mathbf{\tilde{D}\to0}$:

For $\tilde{D}=0$,  $\theta_e=\theta$ [cf. Eq.~(\ref{g21p})]. Assuming $\theta_e=\theta+\epsilon$, where the parameter $\epsilon$ is small compared
to $\theta$, expansion of Eq.~(\ref{g21p}) yields
 \bb
 \epsilon=\frac{2D\cos\theta}{\cos\theta L+\sin\theta D}\,.
 \ee
 Substituting into Eq.~(\ref{fe21p}) gives
 \bb
 \tilde{H}_1^{2-1^+}\sim \left(\frac{1}{2}+\frac{(\pi-2\theta)\sec\theta\csc\theta}{4}\right)\tilde{D}^{-1}\,,\;\; (\theta>0)\,,
 \ee
showing weaker divergence than in the completely wet case, where $\tilde{H}_1^{2-1^+}\sim\tilde{D}^{-2}$.

\item  $\mathbf{\tilde{D}\to1}$:

 At $\tilde{D}=1$, Eq.~(\ref{g21p}) implies $\theta_e=\pi/2+\theta$. We therefore set $\theta_e=\pi/2+\theta-\epsilon$
 with $\epsilon\ll 1$ and expand Eq.~(\ref{g21p}) to first order to get
 \bb
 \epsilon=\frac{(D-L)(\cos\theta-\sin\theta)}{L\sin\theta-D\cos\theta}\,.
 \ee
 Substituting into Eq.~(\ref{fe21p}) yields
%
  \begin{widetext}
  \bb
\tilde{H}_1^{2-1^+}\sim\frac{\pi -4 \theta}{4 \cos^{2}\theta-2}+\frac{\cos \theta \left[4 \cos^{3}\theta-8 \sin  \theta \cos^{2}\theta-2 \cos
 \theta+\left(\pi -4 \theta +4\right)\sin \theta \right]}{{\left(2 \cos^{2}\theta-1\right)}^{2}} \left(1-\tilde{D} \right)\,,\;{\rm
 as}\;\tilde{D}\to1\,.  \label{2as_right_pw}
  \ee
  \end{widetext}
which smoothly reduces to the complete wetting result given by Eq.~(\ref{2as_right_cw}).

\end{enumerate}

\subsubsection{$2$-$1^-$ boundary}
We now turn to the coexistence line between the $2$- and $1^-$- condensations. This occurs when $\theta_e^+=\pi/2+\theta$, so that
Eqs.~(\ref{geom1_2_cw}) and (\ref{geom2_2_cw}) reduce to
 \bb
 R(\cos\theta_e^--\sin\theta)=L\,, \label{21m_g1}
 \ee
 \bb
 R(\cos\theta-\sin\theta_e^-)=D \label{21m_g2}
 \ee
 and hence
 \bb
  \frac{\cos\theta-\sin\theta_e^-}{\cos\theta_e^--\sin\theta}=\frac{D}{L}\,. \label{21m_g}
 \ee

 Inserting $R$ from Eq.~(\ref{21m_g2}) into the free-energy balance, Eq.~(\ref{theta_two_pw}), yields for the $2$-$1^-$ phase boundary:
 \begin{eqnarray}
 &&\tilde{H}_1^{2-1^-}L(\cos\theta-\sin\theta_e^-)^2\nonumber\\
 &&+D^2\left(\frac{\pi}{2}-\theta-\theta_e^--\sin\theta_e^-\cos\theta_e^-+\sin\theta\cos\theta\right)
 \nonumber\\
&& =\cos\theta(\cos\theta-\sin\theta_e^-)D(\tilde{H}_1^{2-1^+-}+D-L)\,, \label{21m_therm}
 \end{eqnarray}
which together with Eq.~(\ref{21m_g}) gives
 \begin{eqnarray}
\tilde{H}_1^{2-1^-}=\frac{2\cos\theta(\cos\theta-s)\tilde{D}(\tilde{D}-1)-A}{(\cos\theta-s)\left(\cos\theta-s-2\tilde{D}\cos\theta\right)}\,.
\label{bound_21m}
 \end{eqnarray}
 Here,
 \bb
A=\tilde{D}^2\left(\frac{\pi}{2}-\theta-\arcsin(s)-s\sqrt{1-s^2}+\sin\theta\cos\theta\right)
 \ee
 and $s\equiv\sin\theta_e^-$ is given by
 \bb
 s=\frac{\cos\theta+2\tilde{D}\sin\theta-\tilde{D}^2\cos\theta}{1+\tilde{D}^2}\,.
 \ee
 as follows from Eq.~(\ref{21m_g}).

The asymptotics of $\tilde{H}_1^{2-1^-}$ in two limiting regimes are:\\

\begin{enumerate}

\item $\mathbf{\tilde{D}\ll1}$:

Setting $\theta_e^-=\pi/2-\theta+\epsilon$ with $\epsilon\ll1$ and expanding Eq.~(\ref{21m_g}) to first order gives
 \bb
  \epsilon=\frac{2\tilde{D}\sin\theta}{D\cos\theta+\sin\theta}\,. \label{eps_21m}
 \ee
 Substituting for $\theta_e^-$ into Eq.~(\ref{21m_therm}) yields
   \begin{widetext}
  \begin{eqnarray}
  \tilde{H}_1^{2-1^-}&=&\frac{\theta  \csc  \theta+\cos \theta}{2 \sin  \theta+2}
  +\frac{\theta\cot  \theta \csc  \theta -\csc  \theta-1 }{\sin\theta+1}\tilde{D}+{\cal{O}}(\tilde{D}^2)\,.
  \end{eqnarray}
  \end{widetext}
Thus, for partially wet walls the $2$-$1^-$ boundary grows linearly with $\tilde{D}$ near the origin, in contrast to the quadratic behaviour for
completely wet walls [cf. Eq.~(\ref{21m_left_cw})].

\item $\mathbf{\tilde{D}\lesssim 1}$:

In this case, Eq.~(\ref{21m_g}) implies that the appropriate expansion is $\theta_e^-=\theta+\epsilon$, with
 \bb
  \epsilon=\frac{(1-\tilde{D})(\cos\theta-\sin\theta)}{\cos\theta-\tilde{D}\cos\theta}\,.
 \ee
Upon substituting for $\theta_e^-$ into Eq.~(\ref{21m_therm}) we obtain
  \begin{eqnarray}
&&\tilde{H}_1=\frac{4 \cos^{2}\theta-4\left(\sin  \theta+1\right) \cos \theta+\pi -4 \theta +4 \sin  \theta}{4\left(\sin \theta+1\right) \cos
\theta-4
\sin \theta-2}\nonumber\\
&& +{\cal{O}}(\tilde{D}-1)\,,
  \end{eqnarray}
  where the ${\cal{O}}(\tilde{D}-1)$ term vanishes for $\theta=0$, recovering (\ref{21m_right_cw}).
Thus, near $\tilde{D}=1$ the boundary exhibits linear behaviour for partially wet walls, again contrasting with the quadratic approach characteristic
of completely wet walls.

\end{enumerate}

\section{Condensation limits}

\begin{figure*}
 \includegraphics[width=\textwidth]{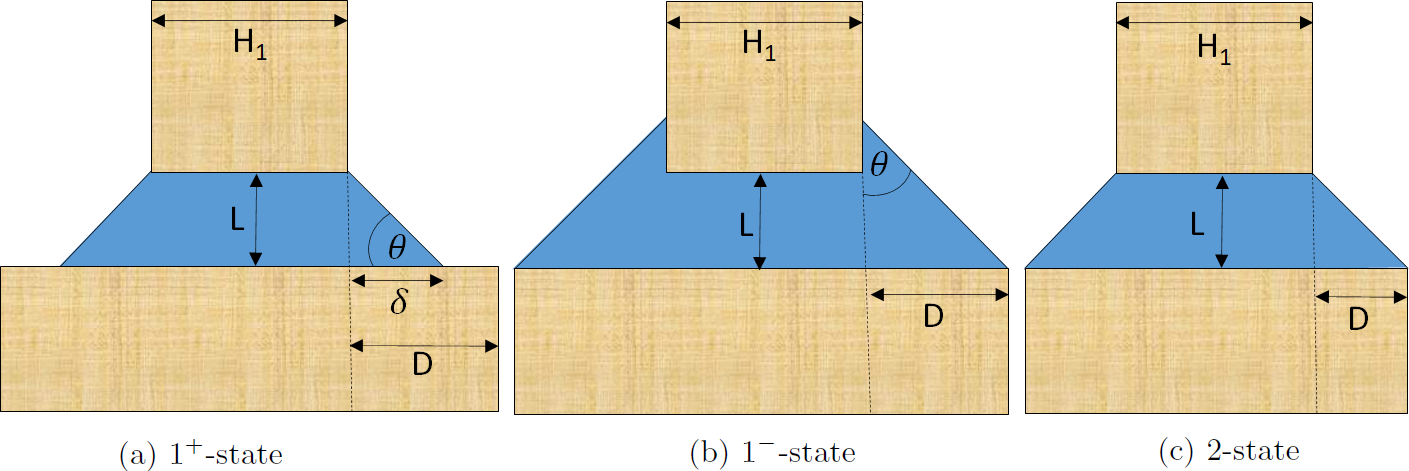}
\caption{Schematic representations of condensation states at saturation: (a) $1^+$, (b) $1^-$, and (c) $2$.} \label{scheme_sat}
\end{figure*}


Each condensation type persists only up to saturation, where the confined system condenses simultaneously with the bulk. For a given $x$-condensation
state, this limit corresponds to $\delta\mu^{(x)}=0$, beyond which the $x$-state ceases to be stable. At saturation the menisci become flat (see
Fig.~\ref{scheme_sat}), which allows for a straightforward geometrical description of the limiting conditions.

In the following, we derive the saturation limits for each condensation type and verify their consistency by independently recovering the limits of
the corresponding phase boundaries. These results provide a consistency check of the preceding analysis, impose additional restrictions on the
occurrence of condensation, and clarify the mechanism by which the topology of the global phase diagrams changes as the contact angle crosses the
wedge-filling threshold $\theta=\pi/4$.

\subsection{$1^+$-condensation}

At saturation, the excess grand potential of the $1^+$-state is (see Fig.~\ref{scheme_sat}a)
 \bb
  \frac{\Omega^{\rm ex}_{1^+}}{2\gamma}=L\csc\theta-(L\cot\theta+H_1)\cos\theta\,.
 \ee
Setting $\Omega^{\rm ex}_{1^+}=0$ gives the saturation limit of $1^+$-condensation as
  \bb
  \tilde{H}_{1m}^{1^+}=\tan\theta\,, \label{1p_limit}
  \ee
  meaning that $1^+$-condensation can only occur when $\tilde{H}_1>\tilde{H}_{1m}^{1^+}$.

This restriction is irrelevant for $\theta<\pi/4$, since both the $1^+$-$2$ and $1^+$-$0$ phase boundaries exist only for $\tilde{H}_1>1$. However,
for $\theta>\pi/4$  condition (\ref{1p_limit}) imposes a strong constraint on the occurrence of $1^+$-condensation and, indeed, of any condensation
at all, as will be discussed later.

Equation~(\ref{1p_limit}) is independent of $D$, provided the condensed region does not extend beyond the bottom wall. This  introduces the
additional restriction $D<\tilde{D}_m^{1^+-2}$, where
 \bb
 \tilde{D}_m^{1^+-2}=\cot\theta \label{1p-2_limit}
 \ee
defines the saturation limit of the $1^+$-$2$ boundary. One can verify that this condition is consistent with the earlier results,
Eqs.~(\ref{bound_21p}) and (\ref{1p_limit}), for $\theta_e=\pi/2-\theta$. No analogous restriction exists for the $1^+$-$0$ line,  since this
boundary is absent for $\theta>\pi/4$.

\subsection{$1^-$-condensation}

The marginal configuration of the $1^-$-state at the highest possible pressure is shown in Fig.~\ref{scheme_sat}b.  Its excess grand potential per
unit length is
 \bb
  \frac{\Omega^{\rm ex}_{1^-}}{2\gamma}=D\csc\theta-(H_1+D-D\cot\theta-L)\cos\theta\,,
 \ee
leading to the minimal wall height required for $1^-$-condensation,
 \bb
  \tilde{H}_{1}^{1^-}=1+\frac{\sin\theta-\cos\theta}{\cos\theta}\tilde{D}\,. \label{1m_limit}
  \ee

  This line is bounded by two characteristic points (cf. Fig.~\ref{fig_pd} below). The first is its intersection with the $2$-$1^-$ phase boundary,
Eq.~(\ref{bound_21m}), which occurs at
 \bb
  \tilde{D}^{1^-}_{\rm min}=\tan\theta\,.  \label{dmin_1m}
 \ee
 The second is the point at which $H_1=0$, given by
  \bb
 \tilde{D}^{1^-}_{0}=\frac{\cos\theta}{\cos\theta-\sin\theta}\,,  \label{d0_1m}
  \ee
  which diverges as $\theta\to\pi/4$.

\subsection{$2$-condensation}

The $2$-state at saturation is shown in Fig.~\ref{scheme_sat}c. The corresponding excess grand potential is
 \bb
  \frac{\Omega^{\rm ex}_{2}}{2\gamma}=\sqrt{L^2+D^2}-(H_1+D)\cos\theta\,,
 \ee
 which vanishes at the saturation limit of $2$-condensation, given by
  \bb
   \tilde{H}_1^{2}=\sqrt{1+\tilde{D}^2}\sec\theta-\tilde{D}\,. \label{2_limit}
  \ee

In the $\tilde{D}$-$\tilde{H}_1$ plane, this curve forms the lower bound of the $2$-condensation region. It extends up to $\tilde{D}=\tan\theta$,
where it meets the $2$-$1^-$ boundary and connects smoothly to the $\tilde{H}_{1}^{1^-}$ line.

\subsection{$0$-condensation}

The only geometrically admissible case for $0$-condensation at saturation occurs when $\theta=\pi/4$, i.e. the wedge-filling threshold, when the
menisci can connect both walls at the correct contact angle. Thermodynamically, however, the precise meniscus position of the menisci is irrelevant:
all $0$-states at $\theta=\pi/4$ are degenerate. Nevertheless, these states remain metastable with respect to the gas, whose free energy is always
lower. This reflects the missing wall segment of length $L$ in our geometry (compared to a true wedge), which prevents the reduction of free energy
by $\gamma L\cos\theta$. Thus, $0$-condensation cannot persist up to saturation.

\subsection{Disappearance of $0$ and $1^-$-condensation at $\frac{\pi}{4}$}

\begin{figure}[bht]
\includegraphics[width=8cm]{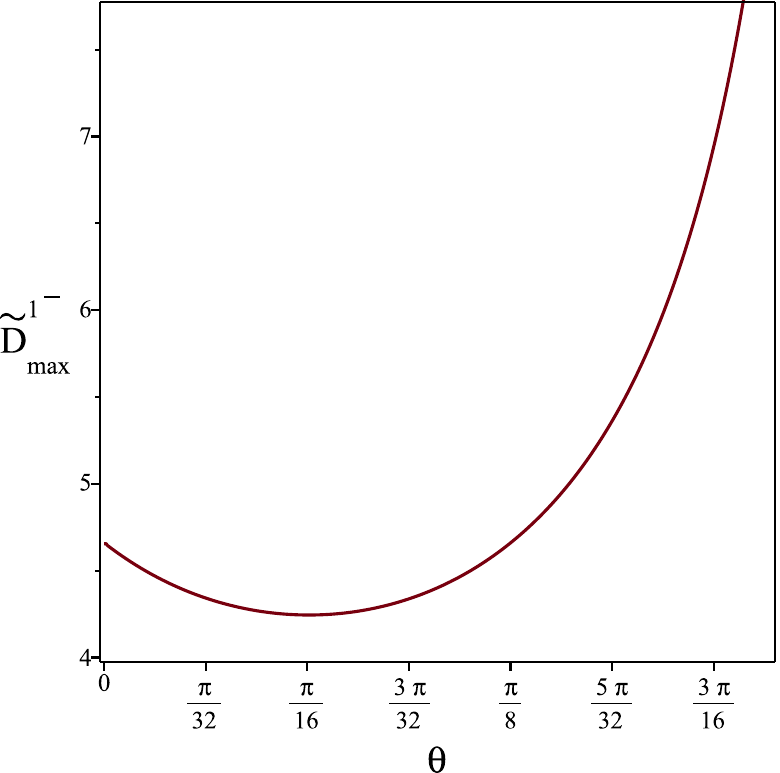}
\caption{Dependence of the maximum extension $\tilde{D}^{1^-}_{\rm max}$ permitting $1^-$-condensation on $\theta$. The curve has a shallow minimum
near $\theta\approx 11^\circ$ and diverges as $\theta\to\pi/4$.} \label{dm}
\end{figure}

It follows directly from geometry that $0$-condensation can only occur for $\theta<\pi/4$ with $D\ge L$. One can ask what is the manner at which this
region vanishes as $\theta\to\pi/4$ in the $\tilde{D}$-$\tilde{H}_1$ diagram. For $\theta<\pi/4$, the $0$ region is bounded above by the $1^+$-$0$
boundary, a semi-infinite horizontal line at $\tilde{D}\ge1$ whose height decreases smoothly from $\pi/2$ to unity. From below, it is bounded by the
$0$-$1^-$ boundary, which intersects the abscissa at $\tilde{D}^{1^-}_{\rm max}$. Interestingly, the dependence $\tilde{D}^{1^-}_{\rm max}(\theta)$
is non-monotonic: it decreases slightly at small $\theta$, reaches a minimum near $11^\circ$, and then diverges as $\theta\to\pi/4$ (see
Fig.~\ref{dm}). This divergence reflects the progressive flattening of the $0$-$1^-$ line, which eventually merges with the $1^+$-$0$ line at
$\theta=\pi/4$. The merging rate is characterized by the slope of $\tilde{H}_1^{0-1^-}$ at $\tilde{D}=1$, which vanishes as
  \bb
   \frac{d\tilde{H}_1^{0-1^-}}{d\theta}\sim\frac{4}{3}\left(\theta-\frac{\pi}{4}\right)\,. \label{dh_01m}
  \ee
Thus, for $\theta<\pi/4$, $0$-condensation persists for all $\tilde{D}\ge1$, but the admissible interval of $\tilde{H}_1$ shrinks continuously with
increasing $\theta$ and disappears completely at $\theta=\pi/4$.

The disappearance of $1^-$-condensation follows a different mechanism. For $\theta<\pi/4$ the $1^-$ region may occur for both $\tilde{D}<1$ and
$\tilde{D}>1$, but both subregions disappear as $\theta\to\pi/4$:

\begin{enumerate}

\item For $\tilde{D}<1$, the $1^-$-region gradually shrinks as the minimum extension $\tilde{D}^{1^-}_{\min}$ along the $2$-$1^-$ line shifts
      toward unity with increasing $\theta$ [cf. Eq.~(\ref{dmin_1m})].

\item For $\tilde{D}>1$, the $1^-$-region is bounded by the $0$-$1^-$ boundary and the $\tilde{H}_{1}^{1^-}$ line.
       As $\theta\to\pi/4$, the latter flattens as
  \bb
 \tilde{H}_{1}^{1^-}\sim 1+2\tilde{D}\left(\theta-\frac{\pi}{4}\right)\,. \label{h_1m}
  \ee
Comparing Eqs.~(\ref{dh_01m}) and (\ref{h_1m}) shows that although $1^-$-condensation always exists for $\tilde{D}>1$ when $\theta<\pi/4$, the
wedge-shaped region between $\tilde{H}_{1}^{0-1^-}$ and $\tilde{H}_{1}^{1^-}$ narrows and closes exactly at $\theta=\pi/4$. For $\theta>\pi/4$, the
slope of $\tilde{H}_{1}^{1^-}$ become positive  and exceeds that of $\tilde{H}_1^{0-1^-}$, which eliminates the $1^-$-region.

Hence, $\theta=\pi/4$ marks the threshold at which $0$- and $1^-$-condensation vanish, leaving only $1^+$- and $2$-condensation as admissible states.
This underlies the topological changes in the global phase diagrams discussed next.

\end{enumerate}

\section{Global phase diagrams}

\begin{figure*}[h]
    \subfloatflex{$\theta=0$}{%
        \includegraphics[width=7.5cm]{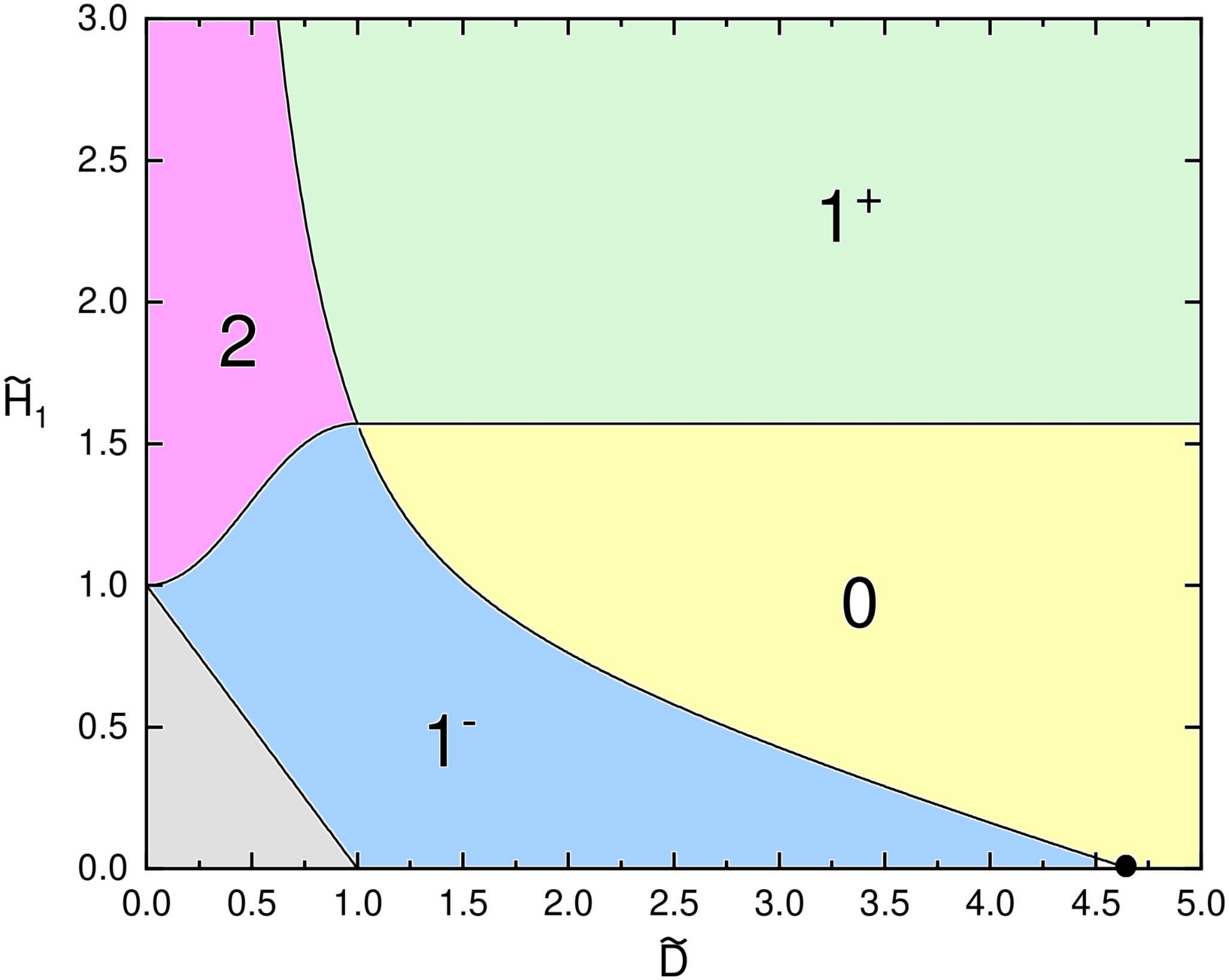}%
    } \hspace*{0.2cm}
    \subfloatflex{$\theta=\pi/8$}{%
        \includegraphics[width=7.5cm]{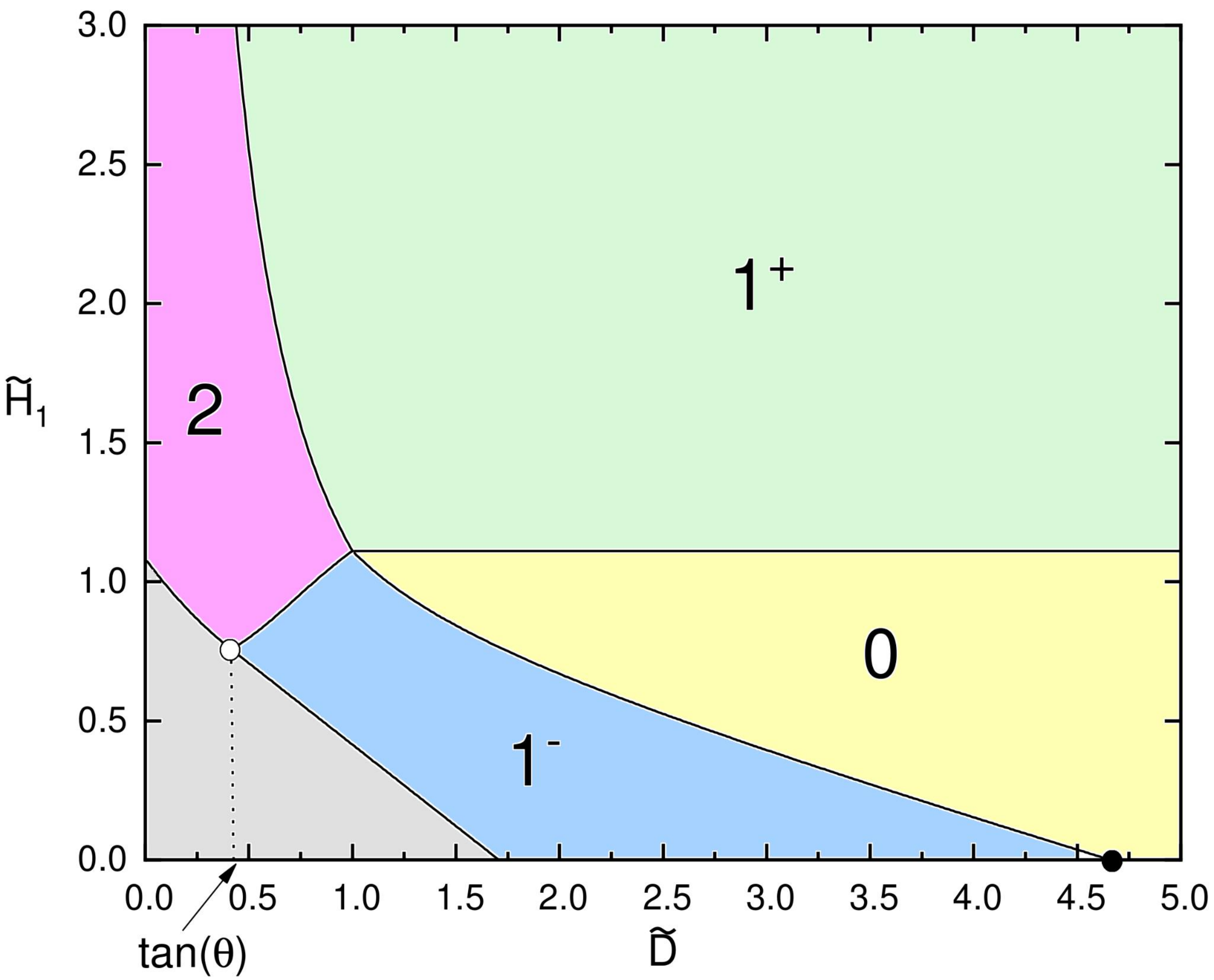}%
    }
    \subfloatflex{$\theta=40\degree$}{%
        \includegraphics[width=7.5cm]{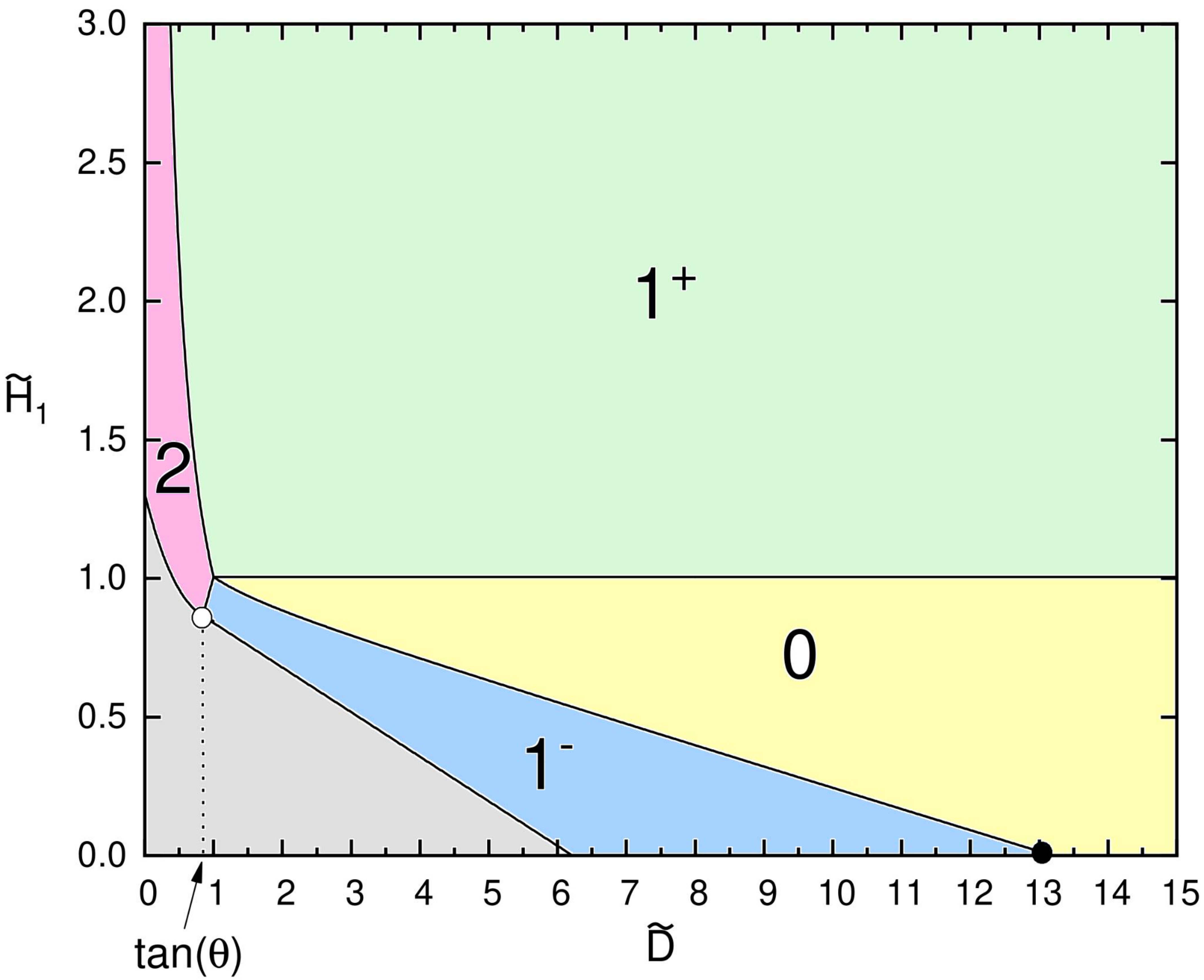}%
    } \hspace*{0.2cm}
    \subfloatflex{$\theta=\pi/4$}{%
        \includegraphics[width=7.5cm]{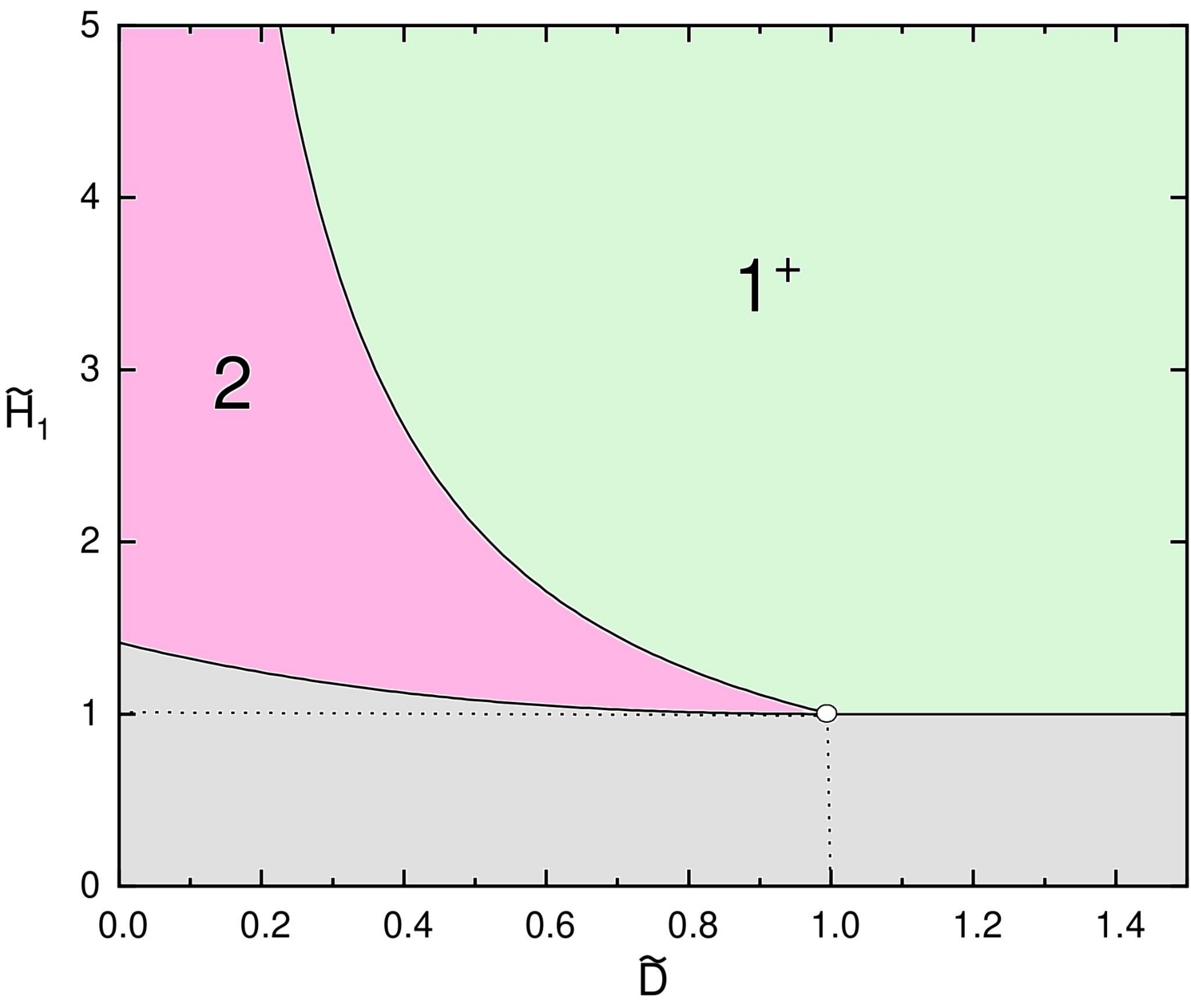}%
    } \hspace*{0.2cm}
    \subfloatflex{$\theta=3\pi/8$}{%
        \includegraphics[width=7.5cm]{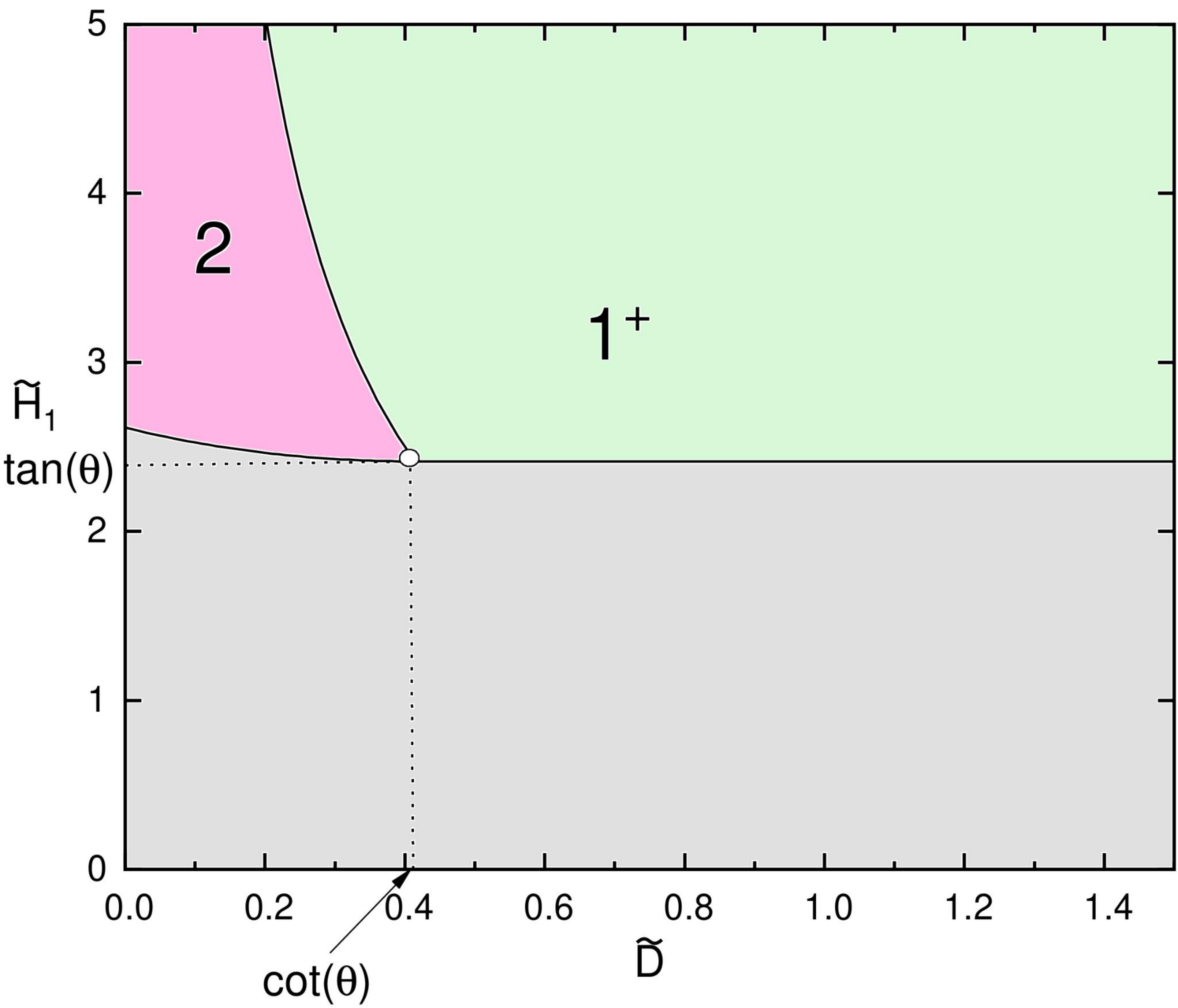}%
    } \hspace*{0.2cm}
    \subfloatflex{$\theta=80\degree$}{%
        \includegraphics[width=7.5cm]{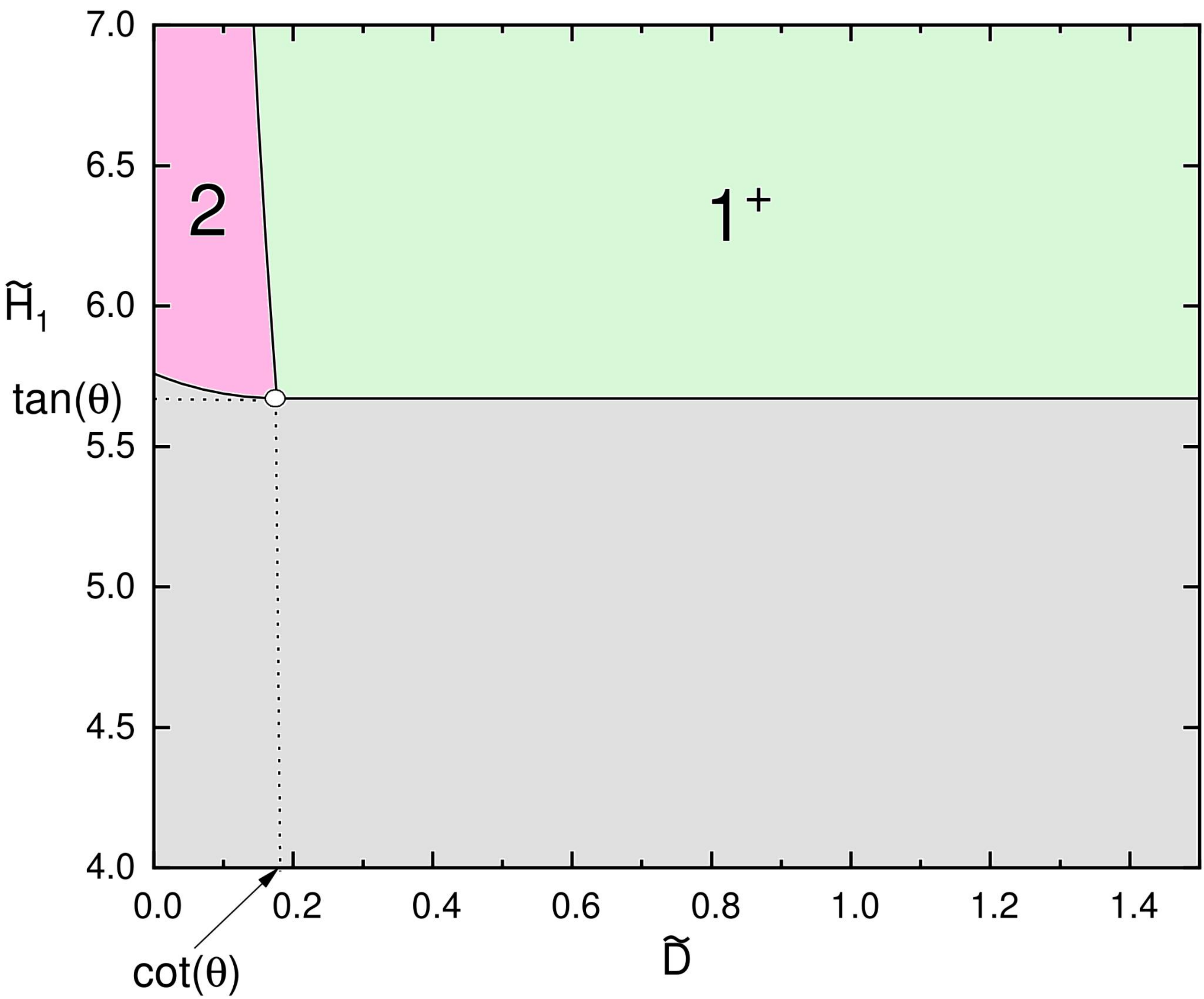}%
    }
    \caption{Global phase diagrams showing condensation types as functions of the reduced parameters $\tilde{D}=D/L$ and $\tilde{H}_1=H_1/L$
        for various contact angles $\theta$. The gray area denotes conditions where no capillary condensation occurs.
        The solid circle marks $\tilde{D}^{1^-}_{\rm max}$, the largest extension admitting $1^-$-condensation.  } \label{fig_pd}
\end{figure*}

In this section, we present global phase diagrams showing the stable types of capillary condensation as functions of system parameters. To this end,
we construct a series of $\tilde{D}$-$\tilde{H}_1$ projections that illustrate how the phase behaviour evolves with increasing contact angle
$\theta$. Particular emphasis is placed on the vicinity of the wedge-filling threshold $\theta=\pi/4$, where the structure of the phase diagrams
changes qualitatively.

We begin with the case of complete wetting, $\theta=0$, where all four condensation types are present (Fig.~\ref{fig_pd}a). Phase boundaries are
shown as solid lines, with $0$ and $1^-$ condensation restricted to $\tilde{H}_1\le\pi/2$. The diagram also contains a triangular region at small
$\tilde{D}$ and $\tilde{H}_1$, corresponding to conditions under which no capillary condensation is possible (hereafter referred to as the no-CC
region). This region connects continuously to the $2$-$1^-$ boundary at $\tilde{D}=0$, indicating that $1^-$-condensation can occur for arbitrarily
small extensions. This feature is unique to $\theta=0$, as is the behaviour of the $2$-$1^+$ line, the only unbounded boundary, which diverges
rapidly as $\tilde{H}_1\sim\tilde{D}^{-2}$ [cf. Eq.~(\ref{2as_left_cw})].

Next, we turn to partially wet walls with contact angles still below the wedge-filling threshold. The case $\theta=\pi/8$ is displayed in
Fig.~\ref{fig_pd}b; the open circle marks the point where the saturation limits $\tilde{H}_{1}^{1^-}$ and $\tilde{H}_{1}^{2}$ meet and cross the
$2$-$1^-$ boundary at $\tilde{D}^{1^-}_{\rm min}=\tan\theta$. As a result, the no-CC region is no longer triangular and extends to larger
$\tilde{D}$. This compresses the $1^-$ region relative to the $\theta=0$ case, while the upper bound $\tilde{D}^{1^-}_{\rm max}$ remains unchanged
for this value of $\theta$ [cf. Fig.~\ref{dm}]. The $2$-condensation region is also reduced, reflecting the slower growth of the $2$-$1^+$ boundary,
which now scales as $\tilde{H}_1^{2-1^+}\sim\tilde{D}^{-1}$. Another qualitative difference relative to the complete wetting case appears in the
$2$-$1^-$ line, which exhibits linear rather than quadratic behaviour at its bounds.

As $\theta$ increases further toward $\pi/4$, the $1^-$ and $0$ condensation types are progressively suppressed. Figure~\ref{fig_pd}c, for
$\theta=40^\circ$, illustrates how $\tilde{D}^{1^-}_{\rm min}$ and $\tilde{H}_1^{0-1^+}$ shift toward unity, while both $\tilde{D}^{1^-}_0$ and
$\tilde{D}^{1^-}_{\rm max}$ grow rapidly. This compresses the $1^-$ and $0$ regions, enlarging the no-CC domain. Right at $\theta=\pi/4$, the $1^-$
and $0$ regions collapse to infinitesimal thickness, reducing to the semi-infinite line $\tilde{H}_1=1$ for $\tilde{D}\ge1$ (Fig.~\ref{fig_pd}d).
Beyond this threshold, condensation requires $H_1\ge1$ and is restricted to $1^+$ and $2$ states.

For larger contact angles, condensation is allowed only under increasingly stringent constraints, requiring $\tilde{H}_1 \geq \tan\theta$, as
illustrated in Fig.~\ref{fig_pd}e for $\theta=3\pi/8$. Here the no-CC region dominates the diagram, while the $2$-condensation is confined to
$\tilde{D}\le\cot\theta$. In the limiting case $\theta\to\pi/2$ (Fig.~\ref{fig_pd}f), the phase diagram is almost entirely governed by $1^+$
condensation, which itself requires gradually larger values of $\tilde{H}_1$.


\section{Conclusion}

We have investigated the phase behaviour of fluids confined between two parallel walls of unequal lengths, $H_1$ and $H_2$, separated by a distance
$L$. Using geometric arguments based on the concept of the edge contact angle, we identified four distinct condensation states, distinguished by how
the menisci  -- separating the condensed liquid from the surrounding bulk gas --  connect to the confining walls: in the $1^+$ state the menisci are
pinned at the edges of the top wall, in the $1^-$ state they are pinned at the edges of the bottom wall, in the $2$ state both walls pin the menisci,
and in the $0$ state the menisci meet the walls at equilibrium contact angles without being pinned.

For each condensation type, we derived Kelvin-like relations specifying the undersaturation at which transitions occur, established the conditions
for their admissibility, and determined the phase boundaries separating them. The phase behaviour is controlled by three dimensionless parameters
$\tilde{H}_1=H_1/L$, $\tilde{D}=D/L$ with $D=(H_2-H_1)/2$, and the Young contact angle $\theta$, which can be tuned by temperature. The analysis
reveals two fundamental geometric constraints that govern the system's phase behaviour: the separatrix at $D=L$, which distinguishes between $0$- and
$2$--condensation, and the wedge-filling threshold at $\theta=\pi/4$, which marks the transition between qualitatively distinct regimes:

\begin{enumerate}

\item $\bm{\theta<\pi/4}$:
All four condensation types are possible. The $2$-state exists only for $\tilde{D}\le1$, while the $0$-state requires $\tilde{D}\ge1$. At
$\tilde{D}=1$, a special value of $\tilde{H}_1$ exists -- ranging from $\pi/2$ (for $\theta=0$) to unity (for $\theta=\pi/4$) -- at which all four
condensation types coexist. For sufficiently large $\tilde{H}_1$, only $1^+$ and $2$ states remain, with their boundary diverging as $\tilde{D}\to0$:
$\sim\tilde{D}^{-1}$ for partially wet walls and $\sim\tilde{D}^{-2}$ for completely wet walls. The phase diagram also contains a compact no-CC
region, where condensation is prohibited; its extent increases with $\theta$ and becomes unbounded at the wedge-filling threshold, $\theta=\pi/4$.

\item$\bm{\theta=\pi/4}$:
At the wedge-filling threshold, the $0$ and $1^-$ regions collapse simultaneously to the semi-infinite line $\tilde{H}_1=1$ with $\tilde{D}\ge1$.
This is the only case where three condensation types coexist along a line.

\item$\bm{\theta>\pi/4}$:
Only $1^+$ and $2$ condensation remain. With increasing $\theta$, the $1^+$ region expands while the $2$ region retreats, confined to
$\tilde{D}\le\cot\theta$. In this regime, condensation requires a minimal upper wall length $\tilde{H}_1=\tan\theta$, which accounts for the complete
disappearance of condensation in the limit $\theta\to\pi/2$.

\end{enumerate}

For $\theta>\pi/2$, the phenomenology reverses: capillary condensation is replaced by capillary evaporation, where the system immersed in bulk liquid
stabilizes a gas-like phase above saturation. Increasing $\theta$ from $\pi/2$ to $\pi$ retraces the same scenarios, but in reverse.



In summary, we have formulated a macroscopic theory of capillary condensation in asymmetric parallel confinements, identifying four distinct
condensation states and deriving the corresponding Kelvin-like equations. The analysis reveals a rich interplay between confinement geometry and
interfacial phenomena, with the wedge-filling threshold $\theta=\pi/4$ and the geometric separatrix at $D=L$ playing the central roles in determining
the phase behaviour.  Beyond their theoretical interest, the present results may also be relevant for experimental or technological situations in
which capillary condensation occurs at finite or truncated boundaries. The different condensation states identified here, together with their
Kelvin-like onset conditions, provide a framework for anticipating how menisci will form and migrate in microfluidic channels, porous or granular
materials, or in micro- and nano-mechanical structures where capillary forces are significant. In such systems, the location and pinning of the
interface often control adhesion, flow resistance, or mechanical stability, and the present analysis illustrates how these factors depend sensitively
on the geometry of the confining walls.

Finally, we comment on several interfacial effects that lie beyond the present macroscopic treatment. Because the confining walls are taken to be
infinitely deep, the liquid-gas interface is always a cylindrical surface of constant mean curvature. Its Gaussian curvature therefore vanishes
identically, and no Gaussian-curvature gradients arise within this geometry; such contributions would appear only if the walls were also finite in
depth. Line tension at the contact line provides an additional boundary term in the free energy and may slightly shift the effective edge contact
angle in very narrow slits. A more fundamental description would also incorporate nonlocal effects, as captured by modern microscopic DFT. These
nonlocal contributions may influence the detailed structure of the meniscus close to edges but do not modify the macroscopic Laplace balance or the
Kelvin-like conditions governing the onset of condensation.

Future work may proceed along several directions. Most importantly, it would be valuable to test the predictions of the present macroscopic theory
against microscopic density-functional calculations or molecular simulations, particularly for narrow slits where packing effects are significant.
Generalizations to heterogeneous systems are also of interest; for instance, walls composed of different materials (and thus exhibiting different
contact angles) would break the symmetry of $0$-condensation and could alter phase behaviour qualitatively \cite{wedge_het}.  Further geometric
modifications may likewise enrich the phenomenology. For example, laterally shifting the two walls to produce unequal overhangs would impose
non-equivalent boundary conditions at the two ends of the slit. In such asymmetric setups, the system could allow for hybrid morphologies in which
distinct meniscus configurations occur at opposite ends, thereby modifying the structure of the phase diagram. Other modifications, such as patterned
or chemically heterogeneous walls, may induce spatially varying edge contact angles and lead to novel pinning-depinning transitions. Similarly,
capping one end of the slit could generate an entirely different phase scenario, potentially changing the nature of the condensations
\cite{our_groove}.  Another natural extension of the present work involves non-parallel or slightly inclined walls, recently analysed in detail in
the context of capillary condensation in open wedges \cite{janek25}. Such inclinations, particularly when combined with finite overhangs, would
introduce additional admissible morphologies and could lead to partially filled or hybrid states that are not captured by the parallel setup
considered here. It would also be interesting to extend the present model to confining walls exhibiting nonzero curvature. Using the method proposed
in Ref. \cite{bridging}, one may be able to describe the corresponding phase behaviour by mapping such systems to an effective finite slit with
unequal overhangs.

Finally, the dynamical aspects of condensation in nanoconfinement remain largely unexplored. Investigating the kinetics of meniscus formation, the
role of  metastable pathways, and the emergence of hysteresis would significantly extend the present work and establish connections to applications
in porous media, microfluidics, or nanofluidics, where precise control of capillary effects is essential.


\begin{thebibliography}{99}

\bibitem{thomson}
W. Thomson, Phil. Mag. {\bf 42}, 448 (1871).

\bibitem{gregg}
S. J. Gregg and K. S. W. Sing, {\it Adsorption, Surface Area and Porosity} (Academic Press, New York, 1982).

\bibitem{row}
J. S. Rowlinson and B. Widom, Molecular Theory of Capillarity (Clarendon, Oxford, 1989).

\bibitem{hansen}
J. P. Hansen and J. R. McDonald, Theory of Simple Liquids (Academic, New York, 2005) 3rd ed.

\bibitem{fisher81}
M. E. Fisher and H. Nakanishi, J. Chem. Phys. {\bf 75}, 5857 (1981).

\bibitem{nakanishi83}
H. Nakanishi and M. E. Fisher, J. Chem. Phys. {\bf 78}, 3279 (1983).

\bibitem{evans84}
R. Evans and P. Tarazona, Phys. Rev. Lett. 52, 557 (1984).

\bibitem{evans86}
 R. Evans, U. Marini Bettolo Marconi, and P. Tarazona, J. Chem. Phys. {\bf 84}, 2376 (1986).

 \bibitem{evans87}
R. Evans and U. Marini Bettolo Marconni, J. Chem. Phys. {\bf 86}, 7138 (1987).


\bibitem{evans79}
R. Evans, Adv. Phys. {\bf 28}, 143 (1979).

 \bibitem{dietrich}
S. Dietrich, in {\it Phase Transitions and Critical Phenomena}, edited by C. Domb and J. L. Lebowitz (Academic, New York, 1988), Vol. 12.

 \bibitem{schick}
M. Schick, in {\it Liquids and Interfaces}, edited by J. Chorvolin, J. F. Joanny, and J. Zinn-Justin (Elsevier, New York, 1990).

\bibitem{derj}
B. V. Derjaguin, Zh. Fiz. Khim. {\bf 14}, 137 (1940).

\bibitem{evans_marc85}
R. Evans and U. Marini Bettolo Marconni, Chem. Phys. Lett. {\bf 114}, 415 (1985).


 \bibitem{tar87}
 P. Tarazona, U. Marini Bettolo Marconi, and R. Evans, Mol. Phys. {\bf 60}, 573 (1987).

 \bibitem{kierlik}
 E. Kierlik and M. Rosinberg, Phys. Rev. A {\bf 44}, 5025 (1991).

 \bibitem{neimark01}
 P. I. Ravikovitch and A. V. Neimark, Colloids Surf. A {\bf 187}, 11 (2001).

 \bibitem{gelb}
 L. D. Gelb, K. E. Gubbins, R. Radhakrishnan, and M. Sliwinska-Bartkowiak, Rep. Prog. Phys. {\bf 62}, 1573 (1999).

 \bibitem{gamble81}
L. R. Fisher, R. A. Gamble, and J. Middlehurst, Nature (London) {\bf 290}, 575 (1981).


 \bibitem{fish_israel}
 L. Fisher and J. Israelachvili, Nature {\bf 277}, 548 (1979).

 \bibitem{hons10}
J. W. van Honschoten, N. Brunets, and N. R. Tas, Chem. Soc. Rev. 39, 1096 (2010).

\bibitem{zhong18}
 J. Zhong, J. Riordon, S. H. Zandavi, Y. Xu, A. H. Persad, F. Mostowfi, and D. Sinton, J. Phys. Chem. Lett. {\bf 9}, 497 (2018).

 \bibitem{geim}
Q. Yang, P. Z. Sun, L. Fumagelli, Y. V. Stebunov, S. J. Haigh, Z. W. Zhou, I. V. Grogorieva, F. C. Wang, and A. K. Geim, Nature {\bf 588}, 250
(2020).


 \bibitem{finite_slit}
A. Malijevsk\'y, A. O. Parry, and M. Posp\'{i}\v sil, Phys. Rev. E {\bf 96}, 020801(R) (2017).

 \bibitem{hauge92}
E. H. Hauge, Phys. Rev. A {\bf 46}, 4994 (1992).

 \bibitem{rejmer99}
K. Rejmer, S. Dietrich, and M. Napi\'orkowski, Phys. Rev. E {\bf 60}, 4027 (1999).

 \bibitem{parry99}
A. O. Parry, C. Rasc\'on, and A. J. Wood, Phys. Rev. Lett. {\bf 83}, 5535 (1999).

 \bibitem{mal13}
A. Malijevsk\'y and A. O. Parry, Phys. Rev. Lett. {\bf 110}, 166101 (2013).

\bibitem{wedge_het}
A. O. Parry, A. Malijevsk\'y and C. Rasc\'on,  Phys. Rev. Lett. {\bf 113}, 146101 (2014).

 \bibitem{our_groove}
A. Malijevsk\'y and A. O. Parry, J. Phys.: Condens. Matter {\bf 26}, 355003 (2014).

 \bibitem{janek25}
J. Janek and A. Malijevsk\'y, Phys. Rev. E {\bf 111}, 045507 (2025).

 \bibitem{bridging}
 A. Malijevsk\'y and  M. Posp\'{i}\v sil, Phys. Rev. E 109, 034801 (2024).



\end{thebibliography}
\end{document}